\documentclass[%
 reprint,
 amsmath,amssymb,
 aps,
prb,
nofootinbib
]{revtex4-2}

\usepackage{physics}
\usepackage{graphicx}
\usepackage{dcolumn}
\usepackage{bm}
\usepackage{siunitx}

\usepackage{mathtools} 

\usepackage{booktabs} 
\usepackage{array}
\usepackage{vtable}

\usepackage{lipsum} 
\usepackage{flushend}  

\usepackage[whole]{bxcjkjatype}
\usepackage{color}

\usepackage{mathrsfs}

\usepackage{txfonts}
\usepackage[bookmarks=false,colorlinks=true,urlcolor=blue,citecolor=blue,linkcolor=blue,breaklinks=true]{hyperref}

\begin{document}

\preprint{APS/123-QED}

\title{Controllable photocurrent generation in Dirac systems with two frequency drives
}

\author{Yuya Ikeda}
\affiliation{%
 Department of Applied Physics, The University of Tokyo, Hongo, Tokyo, 113-8656, Japan
}%

\author{Sota Kitamura}
\affiliation{%
 Department of Applied Physics, The University of Tokyo, Hongo, Tokyo, 113-8656, Japan
}%

\author{Takahiro Morimoto}
\affiliation{%
 Department of Applied Physics, The University of Tokyo, Hongo, Tokyo, 113-8656, Japan
}%

\date{\today}

\begin{abstract}
We study the bulk photovoltaic effect (BPVE) in Dirac and Weyl semimetals under two-frequency light irradiation. We show that the BPVE emerges for centrosymmetric Dirac and Weyl semimetals in the presence of light fields with frequencies $\Omega$ and $2\Omega$. The BPVE under the two frequency drive involves both contributions independent of the carriers lifetime $\tau$ and contributions proportional to $\tau$. 
Our calculations indicate that the photocurrent's direction, magnitude and type can be dynamically controlled by tuning parameters of the driving fields. Furthermore, we find that the tilt of the Dirac cone significantly affects the photocurrent, particularly in mirror symmetry-lacking Weyl semimetals, leading to an anisotropic optical response. These findings provide new insights into the dynamic control of photocurrents in topological semimetals, offering promising applications for optoelectronic devices.
\end{abstract}

\maketitle



\section{INTRODUCTION}

The bulk photovoltaic effect (BPVE)~\cite{Sturman} is a phenomenon where the light irradiation induces photocurrents in the bulk materials. 
The BPVE is of fundamental importance in understanding the optoelectronic behavior of noncentrosymmetric materials, and is also essential for application purposes including  photodetectors and solar cells~\cite{Grinberg2013,Nie2015,tan2016}.
The BPVE is in sharp contrast to conventional photovoltaic effects in p-n junction, where the heterostructure is crucial for dissociating photocarriers with the internal electric field. 
The main mechanisms for the BPVE are  the second-order optical response, which can be divided into the injection current, the shift current, and ballistic current~\cite{dai2022recent, Nagaosa-Morimoto17, Orenstein21}. The injection current arises from the asymmetric photoexcitation in $k$-space, while the shift current is due to the shift of the spatial position of the electron wave packet during photoexcitation.
The ballistic current is induced by scattering of photocarriers with impurities or phonons and usually depends on the relaxation time of photocarriers.
The presence of inversion symmetry $\mathcal{P}$ prohibits the BPVE, while time reversal symmetry $\mathcal{T}$ or $\mathcal{PT}$ symmetry plays an important role in determining which type of photocurrent is generated. 
For instance, the presence of $\mathcal{T}$ symmetry prohibits the linear injection and circular shift responses, while $\mathcal{PT}$ symmetry prohibits the circular injection and linear shift responses~\cite{Ahn20}.

Dirac and Weyl systems have attracted keen attention because of their unique electronic properties characterized by the gapless linear dispersion~\cite{Wehling2014}. These systems are exemplified by 3D Dirac semimetals such as Cd$_3$As$_2$~\cite{Neupane2014CdAs,Liu2014CdAs} and Na$_3$Bi~\cite{Liu2014NaBi}, as well as by 2D materials like graphene~\cite{Novoselov2005}, and surface states of topological insulators~\cite{Zhang2009}. The gapless nature of their energy spectra leads to a variety of intriguing phenomena, including high mobility of charge carriers~\cite{Neupane2014CdAs}, unusual magnetic responses such as chiral anomaly and chiral magnetic effect~\cite{Burkov2015,Morimoto2016CA,Morimoto2016Magneto}. 
Inversion-broken Weyl systems also support nonlinear optical effects such as BPVEs~\cite{Chan17,Matsyshyn21,Ahn20} and quantized circular photogalvanic effect (CPGE) that arises from the Berry curvature of Weyl points~\cite{deJuan17}.
Magnetic Dirac systems, where inversion symmetry $\mathcal{P}$ is broken while $\mathcal{PT}$ is preserved, also support a BPVE~\cite{Ahn20}.

\begin{figure*}[t]
\begin{center}
\includegraphics[width=\linewidth]{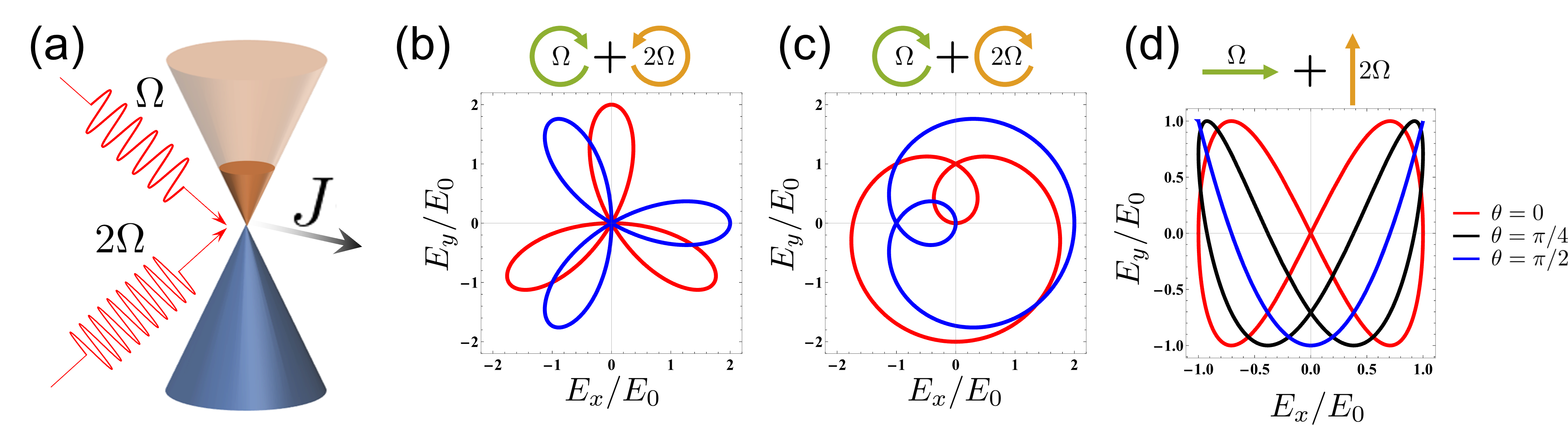}
\caption{
(a) A schematic picture of the third-order dc photocurrent generation in Dirac electron systems.
(b-d) Patterns of electric fields of the two-frequency drive with relative phases $\theta = 0$ (red), $\pi/4$ (black), and $\pi/2$ (blue), for (b) counter-rotating circularly polarized light, (c) co-rotating circularly polarized light, and (d) linearly polarized light.
}
\label{fig_1}
\end{center}
\end{figure*}

The (standard) monochromatic light-induced BPVE manifests only in materials lacking inversion symmetry. If the BPVE could be achieved even in systems with inversion symmetry, photocurrent generation would also occur in centrosymmetric Dirac and Weyl systems, leading to novel optical phenomena. Previous studies have revealed that two-frequency drive, where light of frequencies $\Omega$ and $2\Omega$ is simultaneously incident on a material, can induce third-order photocurrents even in systems with spatial inversion symmetry~\cite{Sipe1996,RIOUX20121,Bas2015,Ikeda23,Dixit24} [See Fig.~\ref{fig_1}(a)].
The two-frequency drive possesses dynamical symmetries that are absent in single-frequency irradiation~\cite{Neufeld2019}.
For example, the counter-rotating bicircular light possesses the dynamical $C_3$ symmetry in its electric field pattern, which is experimentally demonstrated~\cite{Kfir2015,Matsunaga23}.
Such dynamical symmetries of the incident light enables the dynamical control of the symmetry and topology of materials, providing a powerful tool for manipulating electronic and optical properties~\cite{Nag19,Gopal2021,Trevisan22,Sandholzer22,Minguzzi22,Gopal2022,Ikeda2022,Wang2023}. 
Previously, third-order photocurrents have been studied mainly from the perspective of the coherent control induced by quantum interference between one- and two-photon absorption processes~\cite{Sipe1996,Sipe2000,RIOUX20121,Rioux14,Rodrigo14,Zheng22}. 
However, these studies focused mainly on the injection current contribution proportional to the relaxation time $\tau$, leaving the $\tau$-independent contribution induced by two-frequency drive unexplored. 
A comprehensive understanding of the differences between photocurrents generated by monochromatic light and those induced by two-frequency light irradiation is essential.
In particular, a systematic analysis of both the $\tau$-linear and $\tau$-independent contributions in two frequency drives is still missing, and their photocurrent generation in the topological gapless semimetals is an important issue in exploring the future application of two-frequency light.

In this study, we systematically explore both the $\tau$-linear and  $\tau$-independent contributions to the BPVE in Dirac and Weyl systems under two-frequency light irradiation. A key novelty of our work is the comprehensive analysis of the $\tau$-independent contribution, which has not been sufficiently addressed in previous studies. We successfully calculate the conductivity for a range of systems, including 2D massive and massless Dirac, 3D massless Dirac, and tilted Dirac systems. Furthermore, we uncover a significant low-frequency divergence in the photocurrent, which arises due to the gapless dispersion at Dirac/Weyl points. This divergence is stronger than the low-frequency divergence observed in second-order photocurrents of inversion-broken Weyl system. Additionally, we demonstrate that by tuning the polarization of incident laser light, one can dynamically control the generated photocurrent, including directional response separation and control over the photocurrent's type and direction. These findings open up new avenues for utilizing Dirac and Weyl semimetals in tunable optoelectronic applications.

The rest of the paper is organized as follows.
First, we provide a detailed diagrammatic framework for the generation of third-order photocurrents in Dirac and Weyl systems under two-frequency drives in Sec.~\ref{sec2}. 
We derive the analytical expressions for both third-order $\tau$-linear and $\tau$-independent current conductivities in 3D and 2D Dirac/Weyl systems in Sec.~\ref{sec3}, showing that nonmagnetic Dirac and Weyl systems with inversion symmetry $\mathcal{P}$ generally support photocurrents under a two frequency drive. The effect of the tilt of the Dirac cone on the optical third-order conductivities is also discussed analytically.
We analytically demonstrate the components of the photocurrent for pairs of circularly or elliptically polarized lights (bicircular light) and linearly polarized lights (bilinear light) in Sec.~\ref{sec4}. 
We show that by tuning the parameters of the incident light, we can not only control the magnitude and direction of the photocurrent but also manipulate the type of photocurrent and achieve directional response separation. 
Moreover, it is revealed that these photocurrents exhibit low-frequency divergence, which is even more pronounced than the low-frequency divergence in second-order photocurrents in inversion-broken Weyl systems. 
These behaviors are understood by the symmetry consideration and the dimensional analysis in Sec.~\ref{sec5}.
Finally we perform an order estimation of the photocurrent in the Dirac/Weyl materials in Sec.~\ref{sec6}, which indicates  the two frequency drive's potential for high-efficiency optoelectronic applications.

\section{Formalism}\label{sec2}
In this section, we present our setup of the Bloch electrons in solids that are subjected to the  two-frequency driving. 
We then give a description of the diagrammatic framework to study the two-frequency driving.

\subsection{Two-frequency drive}

We consider Bloch electrons in solids that are subjected to the two-frequency drive and are described by the time-dependent Hamiltonian
\begin{align}
    H(t)=H_0(\vb*{k}+e \vb*{A}(t)/\hbar),
\end{align}
where $H_0(\vb*{k})$ is the unperturbed Bloch Hamiltonian in the equilibrium.
Here we write the charge of an electron as $-e$ with $e>0$.
$\vb*{A}(t)$ is the vector potential of the two-frequency drive. The electric field is given by 
\begin{align}
        \vb*{E}(t)&=-\pdv{\vb*{A}(t)}{t}
        \equiv \Re [\vb*{E}^{(\Omega)} e^{i\Omega t}+\vb*{E}^{(-2\Omega)} e^{-2i\Omega t}],
\end{align}
with the complex electric field amplitude $\vb*{E}^{(\Omega)}$ and $\vb*{E}^{(-2\Omega)}$.
The complex field amplitude can take any value; for example, when illuminated on the $x$-$y$ plane, circularly or elliptically polarized light with a ellipticity $\varepsilon$ corresponds to $(E_x^{(\Omega)}, E_y^{(\Omega)})\propto (i,\varepsilon)$, while linearly polarized light (LPL) corresponds to $(E_x^{(\Omega)}, E_y^{(\Omega)})\propto (\cos\phi, \sin\phi)$.
Various two-frequency drive are realized by combining the parameters of the two light waves, $\varepsilon$ or $\phi$.
For example, in the case of the counter-rotating bicircular light, the electric field amplitude can be written as
\begin{align}
    \mqty(E_x(t) \\ E_y(t))=
    E_0
    \mqty(
    -\sin \Omega t +  \sin(2\Omega t -\theta)\\
    \cos \Omega t +  \cos(2\Omega t -\theta)
    ),
\end{align}
where $\theta$  represents the phase difference of the two-frequency drive.
As shown in Fig.~\ref{fig_1}(b), the light pattern draws a rose-like shape with three-fold rotational symmetry.
On the other hand, the co-rotating bicircular light draws a pattern without rotational symmetry as depicted in Fig.~\ref{fig_1}(c).

Not only the type of polarization, but also the phase difference of the two-color beams can significantly alter the patterns of two-frequency fields. 
As shown in Figs.~\ref{fig_1}(b) and (c), for the case of the two-frequency drive consisting of two circularly polarized lights, the phase difference $\theta$ acts like a knob that rotates the field pattern in the two-dimensional plane. On the other hand, for the case of linearly polarized light in Fig.~\ref{fig_1}(d), changing the phase $\theta$ alters the shape of the field pattern itself, rather than just rotating it.  This versatility in controlling the direction or shape of the two-frequency electric field through the relative phase $\theta$ will have important consequences for the photocurrent response, as we will see later.

\subsection{Diagrammatic framework}

\begin{figure}[t]
\begin{center}
\includegraphics[width=\linewidth]{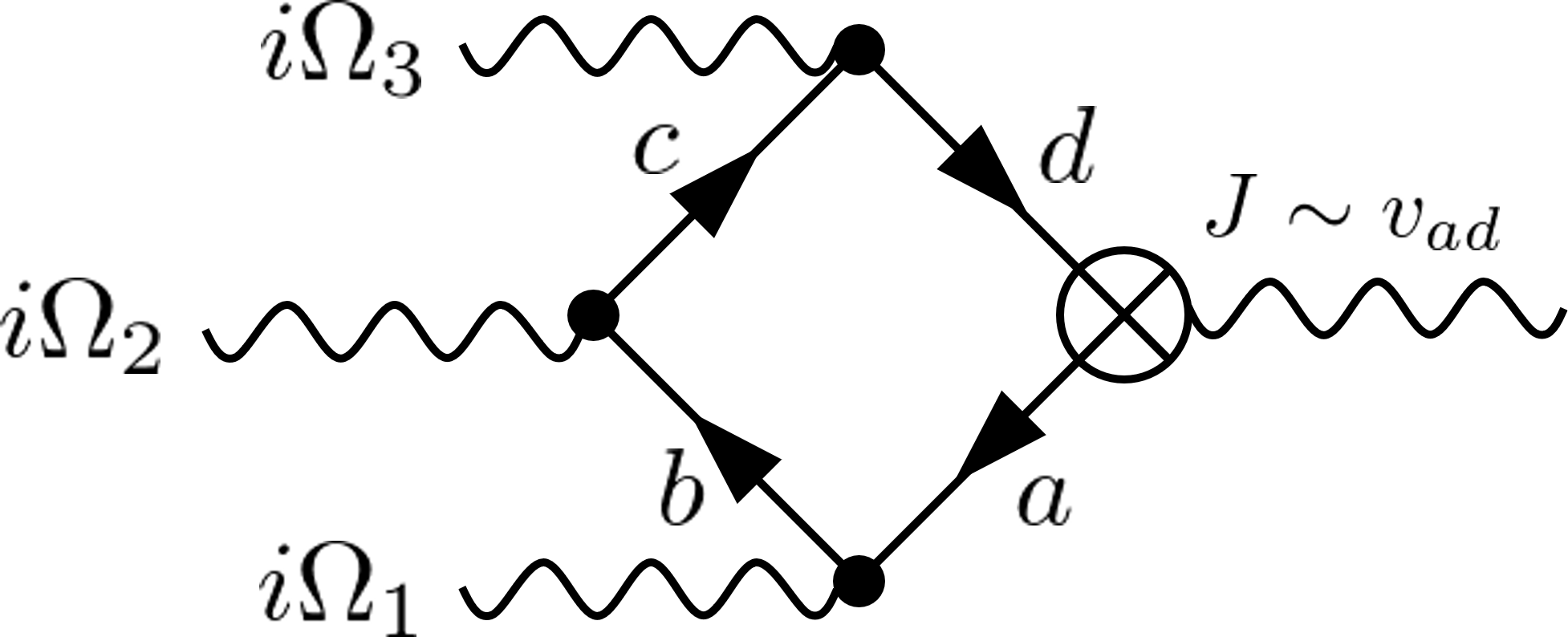}
\caption{The box diagram for the third-order dc generation induced by a two-frequency driving.
The energy of the absorbed photons $\{i\Omega_1, i\Omega_2, i\Omega_3\}$ is given by a permutation of $\{\hbar\Omega+i\gamma_0, \hbar\Omega+i\gamma_0,-2\hbar\Omega+i\gamma_0 \}$.}
\label{fig_diagram}
\end{center}
\end{figure}

In this subsection, we derive the third-order nonlinear dc conductivities defined by
\begin{align}
    J^\mu_{\rm dc} = 
    \sum_{\alpha\beta\gamma} \Re [\sigma^{\mu\alpha\beta\gamma}(0;\Omega,\Omega,-2\Omega) E_\alpha^{(\Omega)} E_\beta^{(\Omega)}E_\gamma^{(-2\Omega)}]
\end{align}
in the Dirac/Weyl electron systems.
Nonlinear optical conductivities can be obtained by the diagrammatic approach \cite{Parker19,Passos2018,YuTzu2024}.
The photocurrent induced by a two-frequency drive is a third-order response with respect to the electric field at the lowest order since $\Omega+\Omega-2\Omega=0$.
Since the Dirac system contains only $k$-linear terms, only the contribution of the box diagram consisting of only one-photon parts remains, as shown in Fig.~\ref{fig_diagram} (for details, see Appendix~\ref{app}).
We note that diagrams including two-photon, three-photon, and four-photon vertices, which correspond to $\partial^2_k H_0$, $\partial^3_k H_0$, and $\partial^4_k H_0$, vanish in the $k$-linear Hamiltonians.
In the box diagram, four black lines represent the electron propagator
$G_a (i\omega)=(i\omega - \epsilon_a)^{-1}$,
where $\epsilon_a$ is the $a$-th original band energy of the system and $i\omega$ is the Matsubara frequency.
Three black vertices stand for the one-photon inputs and we introduce the relaxation rate $\gamma_0$ in the photon energies. The other vertex represents the one-photon output with zero total frequency.
From the Feynman rules, the contribution of the box diagram can be written down as
\begin{align} \label{eq: sigma}
    \sigma^{\mu\alpha\beta\gamma}(0;\Omega,\Omega,-2\Omega)&=
    \frac{-ie^4}{2\hbar^4\Omega^3}\mathcal{S}\sum_{abcd}\int[\dd\vb*{k}]v_{ba}^\alpha v_{cb}^\beta v_{dc}^\gamma v_{ad}^\mu \\ \nonumber
    & \times I_4(\hbar\Omega+i\gamma_0,\hbar\Omega+i\gamma_0,-2\hbar\Omega+i\gamma_0),
\end{align}
where $v_{ba}^\alpha=\bra{b}\pdv{H_0}{k_\alpha}\ket{a}$ represents the matrix element of the velocity operator and $\mathcal{S}$ denotes the summation for all possible permutations of input photons $(\alpha,\Omega),(\beta,\Omega),(\gamma,-2\Omega)$.
Also we defined a shorthand notation for the $\bm k$ integral as $[\dd\vb*{k}]\equiv \dd\vb*{k}/(2\pi)^d$ with the spatial dimension $d$.
The frequency integral in the box diagram, which we call $I_4$, is performed for imaginary-time Green's functions as \cite{Parker19}
\begin{widetext}
\begin{align}
\begin{split}\label{I4}
I_4(i\Omega_1,i\Omega_2,i\Omega_3)&=\int \frac{\dd\omega}{2\pi}    G_a(i\omega)G_b(i\omega+i\Omega_1)G_c(i\omega+i\Omega_1+i\Omega_2)G_d(i\omega+i\Omega_1+i\Omega_2+i\Omega_3)\\
    &=\frac{f(\epsilon_a)}{(\epsilon_{ab}+i\Omega_1)(\epsilon_{ac}+i\Omega_{12})(\epsilon_{ad}+i\Omega_{123})}
        +\frac{f(\epsilon_b)}{(\epsilon_{ba}-i\Omega_1)(\epsilon_{bc}+i\Omega_{2})(\epsilon_{bd}+i\Omega_{23})}\\
        &+\frac{f(\epsilon_c)}{(\epsilon_{ca}-i\Omega_{12})(\epsilon_{cb}-i\Omega_{2})(\epsilon_{cd}+i\Omega_{3})}
        +\frac{f(\epsilon_d)}{(\epsilon_{da}-i\Omega_{123})(\epsilon_{db}-i\Omega_{23})(\epsilon_{dc}-i\Omega_{3})},
\end{split}    
\end{align}
\end{widetext}
where $\epsilon_{ab}=\epsilon_a-\epsilon_b$, $i\Omega_{12}=i\Omega_1+i\Omega_2$, and $i\Omega_{123}=i\Omega_1+i\Omega_2+i\Omega_3$.
$I_4$ in Eq.~\eqref{eq: sigma} is obtained after analytic continuation of Matsubara frequencies,
\begin{align}
    i\Omega_1 &\to \hbar\Omega+i\gamma_0, &
    i\Omega_2 &\to \hbar\Omega+i\gamma_0, &
    i\Omega_3 &\to -2\hbar\Omega+i\gamma_0,
\end{align}
in the above expression, where $\gamma_0$ corresponds to the energy broadening.
Note that the Matsubara frequency $i\omega$ is defined such that it has the dimension of the energy.

It is known that the second-order dc generation in non-centrosymmetric materials arises from two mechanisms in the clean limit: the injection current and the shift current.
The shift current occurs due to changes of 
the real space position of the electron during the optical excitation. 
The injection current is a relaxation process and is proportional to the relaxation time $\tau=\hbar/\gamma_0$, while the shift current is independent of the relaxation time.
By focusing on interband transitions in the clean limit, the third-order dc generation can also be divided into two components: the contribution proportional to the relaxation time $\tau$ and the one independent of $\tau$ in a similar manner to the second order responses. 
Specifically, we decompose the third order nonlinear conductivity as
\begin{align}
    \sigma_{\rm dc}=\sigma_1+ \sigma_0 + \mathcal{O}(\gamma_0) \quad (\gamma_0\to 0),\label{eq:cond-expand}
\end{align}
with $\sigma_{\rm dc}=\sigma(0;\Omega,\Omega,-2\Omega)$,
where the two conductivities $\sigma_i \propto \tau^i$ are defined as
\begin{align}
    \sigma_1&=\frac{1}{\gamma_0}
    \lim_{\gamma_0\to 0}\gamma_0 \sigma_{\rm dc}, \\
    \sigma_0&=\lim_{\gamma_0\to 0}(\sigma_{\rm dc}-\sigma_1).
\end{align}
From the above expressions, we can obtain the following general formulas for two-band models, for example,
\begin{align}
    \sigma_1^{xxxx}&=\frac{-2i e^4}{3\gamma_0 \hbar^5 \Omega^4}\int [\dd \vb*{k}] |v^x_{12}|^2 (\Delta^x)^2   \nonumber \\ 
    &\hspace{2.0cm} \times\qty(\frac{1}{2}\delta(\epsilon_{12}+\hbar\Omega)-\delta(\epsilon_{12}+2\hbar\Omega)), \\ \nonumber
    \sigma_0^{xxxx}&=\frac{-e^4}{3\hbar^6 \Omega^5}\int [\dd \vb*{k}] |v^x_{12}|^2 \Bigl[24|v^x_{12}|^2-(\Delta^x)^2 \Bigr]\\
    &\hspace{1.8cm} \times \qty(\frac{1}{2}\delta(\epsilon_{12}+\hbar\Omega)-\delta(\epsilon_{12}+2\hbar\Omega)),
\end{align}
where $\Delta^x=v^x_{11}-v^x_{22}$ is the difference of the velocity between the valence and conduction band.
We can see that the photocurrent induced by two-frequency driving involves two interband resonance terms: the $\Omega$-resonant term
$\propto \delta(\epsilon_{12}+\hbar\Omega)$ and 
the $2\Omega$-resonant term
$\propto\delta(\epsilon_{12}+2\hbar\Omega)$ with a sign change.

\section{OPTICAL CONDUCTIVITY IN DIRAC SYSTEMS}\label{sec3}
\subsection{3D Dirac/Weyl electrons}
We focus on the photocurrent generation induced by two-frequency drives for Dirac/Weyl semimetals, which are characterized by their gapless linear dispersion. We first consider a 3D model representing a Weyl fermion with an anisotropy in velocity along the $z$ direction, which is given by the Hamiltonian,
\begin{align}\label{ham_3D}
    H^{\rm 3D}= \hbar v_F (k_x \sigma_x + k_y \sigma_y + \eta k_z \sigma_z).
\end{align}
Here $v_F$ is a Fermi velocity and $\eta$ denotes the anisotropy along the $z$ direction.

Let us consider the third-order dc conductivities in this 3D Dirac model.
Because of $C_2^i\ (i=x,y,z)$ symmetry and the ${\rm SO}(2)$ rotational
symmetry around the $z$-axis in the Weyl Hamiltonian as in Eq.~(\ref{ham_3D}), there are only eight independent components,
\begin{align}
\begin{split} \label{eight_cond}
    \sigma^{xxxx},\quad \sigma^{xxzz},\quad \sigma^{xzxz},\quad \sigma^{xzzx},\\
    \sigma^{zzzz},\quad \sigma^{zzxx},\quad \sigma^{zxzx},\quad \sigma^{zxxz}.
\end{split}
\end{align}
Since ${\rm SO}(2)$ symmetry around the $z$-axis is imposed, only two-dimensional components remain, and three-dimensional components such as $\sigma^{xyzz}$ disappear.
The other components obtained by replacing $x$ with $y$ in Eq.~(\ref{eight_cond}) are the same as before the replacement; e.g., $\sigma^{yyzz}=\sigma^{xxzz}$. The components obtained by replacing $z$ with $y$ in Eq.~(\ref{eight_cond}) can be obtained by substituting $\eta = 1$.
We note that the presence of a tilt of the Dirac cone lowers the symmetry of the Dirac cone and leads to additional components, which we discuss in Sec.\ref{sec_tilt}.

By incorporating a variable change as $(k_x,k_y,k_z)=k(\sin\varphi \cos\psi,\sin\varphi \sin\psi, \cos\varphi /\eta)$, we can perform the $k$-integration analytically to compute the $\tau$-linear and $\tau$-independent conductivities as
\begin{align}\label{sigma_3D_inj}
    \mqty(
\sigma_1^{xxxx}\\
\sigma_1^{xxzz}\\
\sigma_1^{xzxz}\\
\sigma_1^{xzzx}
    )&=\frac{-4ie^4 |v_F|\Theta_0}{45\pi \gamma_0\hbar^2\Omega^2 |\eta|}
    \mqty(
    1 \\ 2\eta^2 \\ 2\eta^2 \\ -3\eta^2
    ), \quad
    \mqty(
\sigma_1^{zzzz}\\
\sigma_1^{zzxx}\\
\sigma_1^{zxzx}\\
\sigma_1^{zxxz}
    )=\frac{-4ie^4 |v_F|\Theta_0}{45\pi \gamma_0\hbar^2\Omega^2 |\eta|}
    \mqty(
    \eta^4 \\ 2\eta^2 \\ 2\eta^2 \\ -3\eta^2
    ),\\ \label{sigma_3D_shift}
    \mqty(
\sigma_0^{xxxx}\\
\sigma_0^{xxzz}\\
\sigma_0^{xzxz}\\
\sigma_0^{xzzx}
    )&=\frac{-2e^4 |v_F|\Theta_0}{45\pi \hbar^3\Omega^3 |\eta|}
    \mqty(
    23 \\ 16\eta^2 \\ 16\eta^2 \\ -9\eta^2
    ), \quad
    \mqty(
\sigma_0^{zzzz}\\
\sigma_0^{zzxx}\\
\sigma_0^{zxzx}\\
\sigma_0^{zxxz}
    )=\frac{-2e^4 |v_F|\Theta_0}{45\pi \hbar^3\Omega^3 |\eta|}
    \mqty(
    23\eta^4 \\ 16\eta^2 \\ 16\eta^2 \\ -9\eta^2
    ).
\end{align}
Here the symbol $\Theta_0=\Theta(\hbar \Omega-2\epsilon_F)/8-\Theta(\hbar \Omega-\epsilon_F)$ with the step function $\Theta(x)$ represents the condition for the optical excitation within Dirac cones.
This shows that the $\tau$-linear conductivities are purely imaginary, whereas the $\tau$-independent conductivities are real.
This distinction is attributed to the difference in whether $3i\gamma_0$ is picked up from the energy denominator $I_4$ [see Eq.~(\ref{I4})].
It is worth emphasizing again that the generation of third-order photocurrent driven by a two-frequency field does not require the breaking of spatial inversion symmetry. Especially, the optical conductivities given by Eqs.~(\ref{sigma_3D_inj},\ref{sigma_3D_shift}) are even with respect to both $v_F$ and $\eta$, indicating that when considering a Weyl point pair connected by spatial inversion symmetry or mirror symmetry, the photocurrent do not cancel out but rather simply double.

\begin{figure*}[t]
\begin{center}
\includegraphics[width=\linewidth]{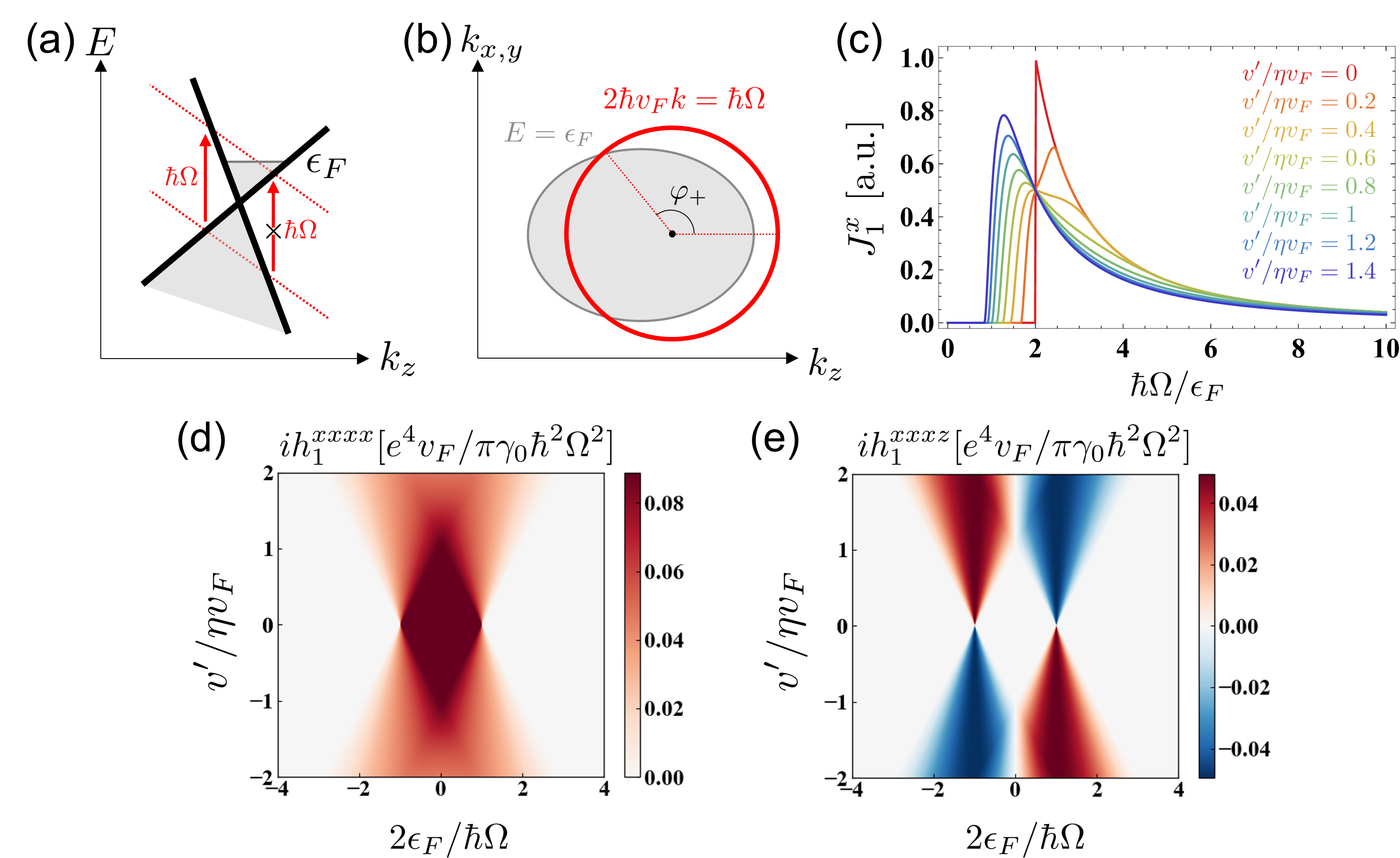}
\caption{
Third order dc conductivities in tilted 3D Dirac electron systems. 
(a) Schematic illustration of the anisotropic optical transitions in a tilted Dirac cone system.
(b) The red circle represents the surface of the resonance condition $\epsilon_{12}=\hbar\Omega$. The gray area shows the Fermi surface. The Fermi surface is distorted by the tilt, and the optical transition can occur only for a part of the red circle outside the gray area due to the Pauli blocking.
(c) Frequency dependence of the photocurrent for various tilt strengths with $\eta=1/4$. As the tilt increases, the peak shifts towards lower frequencies as optical transitions become possible even for $\hbar\omega < 2\epsilon_F$.
(d,e) Density plots of the function $h=\mathcal{F}(\varphi_+)-\mathcal{F}(\varphi_-)$, which characterizes the photocurrent, as a function of tilt strength and frequency. Panel (d) shows the term $ih_1^{xxxx}$ that exists even without tilt, while panel (e) shows the term $ih_1^{xxxz}$ induced by the tilt.
}
\label{fig_tilt}
\end{center}
\end{figure*}

\subsection{2D Dirac electrons}

Next, we turn our focus toward two-dimensional Dirac electron systems represented by
\begin{align}
    H^{\rm 2D}= \hbar v_F (k_x \sigma_x + \eta k_y \sigma_y) + m \sigma_z,
\end{align}
This Hamiltonian describes, for example, the Dirac surface states of a topological insulator or the Dirac fermions in graphene with valley and spin degeneracy.
In the massless case $(m=0)$, we obtain
\begin{align} \label{cond_2Dmassless_inj}
    \mqty(
\sigma_1^{xxxx}\\
\sigma_1^{xxyy}\\
\sigma_1^{xyxy}\\
\sigma_1^{xyyx}
    )&=\frac{-ie^4 v_F^2\Theta_0'}{12 \gamma_0\hbar^2\Omega^3 |\eta|}
    \mqty(
    1 \\ 3\eta^2 \\ 3\eta^2 \\ -5\eta^2
    ), \quad
    \mqty(
\sigma_1^{yyyy}\\
\sigma_1^{yyxx}\\
\sigma_1^{yxyx}\\
\sigma_1^{yxxy}
    )=\frac{-ie^4 v_F^2\Theta_0'}{12\gamma_0 \hbar^2\Omega^3 |\eta|}
    \mqty(
    \eta^4 \\ 3\eta^2 \\ 3\eta^2 \\ -5\eta^2
    ),\\ \label{cond_2Dmassless_shi}
    \mqty(
\sigma_0^{xxxx}\\
\sigma_0^{xxyy}\\
\sigma_0^{xyxy}\\
\sigma_0^{xyyx}
    )&=\frac{-e^4 v_F^2 \Theta_0'}{24 \hbar^3\Omega^4 |\eta|}
    \mqty(
    17 \\ 3\eta^2 \\ 3\eta^2 \\ 11\eta^2
    ), \quad
    \mqty(
\sigma_0^{yyyy}\\
\sigma_0^{yyxx}\\
\sigma_0^{yxyx}\\
\sigma_0^{yxxy}
    )=\frac{-e^4 v_F^2 \Theta_0'}{24 \hbar^3\Omega^4 |\eta|}
    \mqty(
    17\eta^4 \\ 3\eta^2 \\ 3\eta^2 \\ 11\eta^2
    ),
\end{align}
where $\Theta_0'=\Theta(\hbar \Omega-2\epsilon_F)/4-\Theta(\hbar \Omega-\epsilon_F)$. The other components vanish identically. 
As in the 3D massless Dirac case, eight nonzero conductivities have been identified from symmetry considerations.

For the massive case $(m \neq 0)$, we find the effect of the mass gap on the optical conductivities, 
\begin{align}\label{cond_2Dmassive}
    \sigma_1^{xxxx}
    =\frac{-ie^4 v_F^2}{12 \gamma_0\hbar^2\Omega^3 |\eta|}
    \qty[\frac{1}{4}
    g^{xxxx}_1\qty(\frac{2m}{\hbar \Omega},\frac{2\epsilon_F}{\hbar \Omega})-g^{xxxx}_1\qty(\frac{m}{\hbar \Omega},\frac{\epsilon_F}{\hbar \Omega})
    ],
\end{align}
where
\begin{align}
    g^{xxxx}_1(x,y)=\Theta(1-x)\Theta(1-y) (1+2x^2-3x^4) 
\end{align}
is a function that takes the ratio of the input frequency to the mass and Fermi level as arguments.
$g^{xxxx}_1$ contains essential information on the conditions for optical excitations, and is always positive.
One can immediately see that this is consistent with the case $m = 0$.
Due to the breaking of mirror symmetries $M_x,M_y$ induced by the mass term $m$, the other conductivity components, e.g. $\sigma^{xxxy}$, can be finite as
\begin{align}\label{cond_2Dmassive-2}
    \sigma_1^{xxxy}
    =\frac{2e^4 v_F^2 {\rm sgn}(\eta)}{3 \gamma_0\hbar^2\Omega^3}
    \qty[\frac{1}{4}
    g^{xxxy}_1\qty(\frac{2m}{\hbar \Omega},\frac{2\epsilon_F}{\hbar \Omega})-g^{xxxy}_1\qty(\frac{m}{\hbar \Omega},\frac{\epsilon_F}{\hbar \Omega})
    ],
\end{align}
with $g^{xxxy}_1(x,y)=\Theta(1-x)\Theta(1-y) (x-x^3)$.
The $m$-induced $\tau$-linear ($\tau$-independent) conductivities are real (purely imaginary), while the other $\tau$-linear ($\tau$-independent) conductivities are purely imaginary (real).
In addition, these conductivities are characterized by dependence on the chirality of the Dirac cone, i.e. ${\rm sgn}(\eta)$.

\subsection{Effect of the tilt}\label{sec_tilt}
In this section, we discuss the effect of tilt on the dc conductivities.
The tilt of the dispersion causes optical excitation to be anisotropic around the Dirac point in the $k$-space due to the Pauli blocking (rather than occurring on the full sphere around the Dirac point) as shown schematically in Figs.~\ref{fig_tilt}(a,b), whose consequence was discussed in the case of second order optical conductivity \cite{deJuan17,Chan17}.
Here we consider a 3D Dirac electron with a tilt along the $z$-axis that is described by
\begin{align}\label{ham_3D_tilt}
    H^{\rm 3D}_{\rm tilt}= \hbar v' k_z \sigma_0 +\hbar v_F (k_x \sigma_x + k_y \sigma_y + \eta k_z \sigma_z),
\end{align}
where $v'$ represents the overall velocity shift due to the tilt. 
The Dirac fermion described by this Hamiltonian is called type-I when $|v'/v_F|<1$
and type-II when $|v'/v_F|>1$ \cite{Soluyanov2015,Xu2015Weyl}.

The dc conductivities can be expressed in the form
\begin{align}
    \sigma^{\mu\alpha\beta\gamma} &= 
    \frac{1}{8}h^{\mu\alpha\beta\gamma}\qty(\frac{v'}{\eta v_F}, \frac{2\epsilon_F}{\hbar\Omega}) - h^{\mu\alpha\beta\gamma}\qty(\frac{v'}{\eta v_F}, \frac{\epsilon_F}{\hbar\Omega})
    ,\\
    h^{\mu\alpha\beta\gamma}(x,y) &= \mathcal{F}^{\mu\alpha\beta\gamma}(\varphi_+)-\mathcal{F}^{\mu\alpha\beta\gamma}(\varphi_-).
\end{align}
Here, the function $\mathcal{F}$ is obtained as an indefinite integral with respect to $\varphi$ in the $k$-integration Eq.~(\ref{eq: sigma}).
For example, the component $\mathcal{F}_1^{xxxx}$ [defined by $\mathcal{F}=\mathcal{F}_1+\mathcal{F}_0+\mathcal{O}(\gamma_0)$ as in Eq.~(\ref{eq:cond-expand})] is given by 
\begin{align}
    \mathcal{F}_1^{xxxx}(\varphi) =
    \frac{-ie^4 |v_F|}{5760 \pi \gamma_0\hbar^2\Omega^2 |\eta|}
    \qty( -270\cos\varphi +5\cos3\varphi+9\cos5\varphi ).
\end{align}
$\varphi_+$ and $\varphi_-$ are the maximum and minimum $\varphi$ that satisfies 
\begin{align} \label{ineq_phi}
 |x\cos\varphi-y|<1,
\end{align}
which means the condition that only the lower band is occupied and optical excitation occurs with $(k_x,k_y,k_z)=k(\sin\varphi \cos\psi,\sin\varphi \sin\psi, \cos\varphi /\eta)$ [See Fig.~\ref{fig_tilt}(b)].
Note that the dc conductivities without the tilt [Eq.~(\ref{sigma_3D_inj})] can be recovered by substituting $\varphi_+=\pi$ and $\varphi_-=0$ into the above expression.
Introducing a nonzero tilt for the Dirac cone makes the optical transitions anisotropic in the $k$ space and broadens the frequency range of the optical transition, where the lower bound of the allowed frequency for the optical transition is given by
\begin{align}
    \hbar \Omega_{\min} = \frac{2\epsilon_F}{1+ v'/ \eta 
    v_F}.
\end{align}
This leads to a shift of the frequency peak towards lower frequencies, as shown in Fig~\ref{fig_tilt}(c).

The presence of the tilt breaks the mirror symmetry $M_z$, and allows the emergence of additional components other than the eight conductivity tensor components listed in Eq.~(\ref{eight_cond}).
While the remaining SO(2) symmetry around the $z$-axis prohibits the three dimensional components, the following eight tensor components appear due to the tilt as
\begin{align}
\begin{split} \label{add_cond}
    \sigma^{xxxz},\quad \sigma^{xxzx},\quad \sigma^{xzxx},\quad \sigma^{xzzz},\\
    \sigma^{zzzx},\quad \sigma^{zzxz},\quad \sigma^{zxzz},\quad \sigma^{zxxx}.
\end{split}
\end{align}
For example, we obtain
\begin{align}
    \mathcal{F}_1^{xxxz}(\varphi) =
    \frac{-2ie^4 |v_F| {\rm sgn}(\eta)}{45 \pi^2 \gamma_0\hbar^2\Omega^2}
    \qty(-6+\cos2\varphi ) \sin^3\varphi.   
\end{align}
Again, the dc conductivity at zero tilting vanishes because $\mathcal{F}_1^{xxxz}(\pi)-\mathcal{F}_1^{xxxz}(0)=0$.
Unlike the massive 2D Dirac case, $\mathcal{F}_1^{xxxx}$ and $\mathcal{F}_1^{xxxz}$ are both purely imaginary because the additional components induced by the tilt originate from anisotropic optical excitations in $k$-space.

As representative 3D Dirac semimetals, we can consider Dirac materials such as Cd$_3$As$_2$ and Na$_3$Bi, which host two Dirac cones~\cite{Neupane2014CdAs,Liu2014CdAs,Liu2014NaBi}. These two cones are connected by the mirror symmetry $M_z$ of the crystal, and they possess opposite signs of tilt.
The reversed tilt slope, $v'\to -v'$, corresponds to the angle transformation $(\varphi_+,\varphi_-)\to (\pi-\varphi_-,\pi-\varphi_+)$, which is equivalent to reversing the Fermi level $\epsilon_F \to -\epsilon_F$ as seen from Eq.~(\ref{ineq_phi}).
The original tensor components [Eq.~(\ref{eight_cond})] satisfy $\mathcal{F}(\pi-\varphi)=-\mathcal{F}(\varphi)$, whereas the additional terms induced by the tilt [Eq.~(\ref{add_cond})] satisfy $\mathcal{F}(\pi-\varphi)=\mathcal{F}(\varphi)$.
Consequently, the original components fulfill the relationship
$h^{\mu\alpha\beta\gamma}(-x,y)=h^{\mu\alpha\beta\gamma}(x,-y)=h^{\mu\alpha\beta\gamma}(x,y)$, while the additional components fulfill $h^{\mu\alpha\beta\gamma}(-x,y)=h^{\mu\alpha\beta\gamma}(x,-y)=-h^{\mu\alpha\beta\gamma}(x,y)$
[see Figs.~\ref{fig_tilt}(d,e)].
Therefore, in tilted Dirac systems where the crystal hosts mirror symmetry as a whole, the additional components induced by the tilt turn out to cancel out for two Dirac cones. This is also understandable from the symmetry and tensor component arguments discussed previously.
However, if one examines lower symmetric Weyl systems \cite{Hirayama15,Flicker18}, it will become possible to measure the tilt-induced contributions to the photocurrent induced by the two-frequency drive.

\section{PHOTOCURRENT UNDER TWO-FREQUENCY DRIVE}\label{sec4}
Since the third-order nonlinear dc conductivities are derived analytically in the previous section, we obtain the photocurrent simply by multiplying the complex amplitude of the two-frequency drive as
\begin{align}
   J_1^\mu(\Omega)&=
    \sum_{\alpha\beta\gamma} \Re [\sigma_1^{\mu\alpha\beta\gamma} E_\alpha^{(\Omega)} E_\beta^{(\Omega)}E_\gamma^{(-2\Omega)}], \\
    J_0^\mu(\Omega)&=\sum_{\alpha\beta\gamma} \Re [\sigma_0^{\mu\alpha\beta\gamma} E_\alpha^{(\Omega)} E_\beta^{(\Omega)}E_\gamma^{(-2\Omega)}].
\end{align}
In this section, we demonstrate how we can dynamically control the photocurrent by tuning the parameters of the incident light.

\subsection{Circular polarization}\label{sec_circular}

\begin{figure*}[t]
\begin{center}
\includegraphics[width=\linewidth]{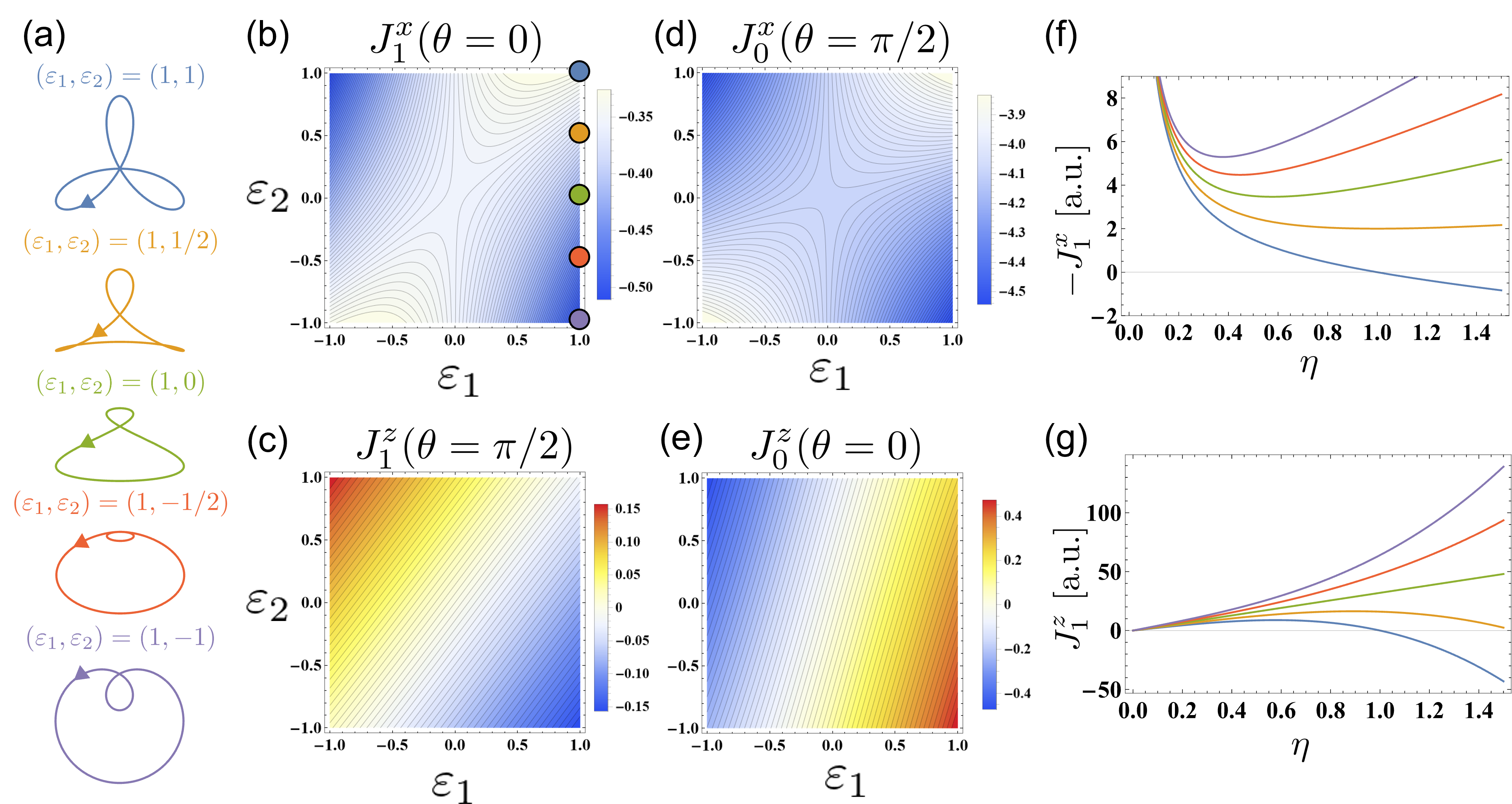}
\caption{
Photocurrent in 3D Dirac electron systems induced by circularly polarized two-frequency electric fields.
(a) Patterns of the circularly polarized two-frequency electric field with different ellipticities $(\varepsilon_1,\varepsilon_2)=(1,1),(1,1/2),(1,0),(1,-1/2),(1,-1)$ and $\theta$ is fixed to 0.
(b-e) Color plots of the photocurrent components as a function of the ellipticities $\varepsilon_1$ and $\varepsilon_2$. The dots in (b) correspond to the patterns of the two-frequency drive shown in (a) with matching colors. The calculations were performed with $\eta = 1/4$, $\theta=0$ in (b,e) and $\theta=\pi/2$ in (c,d). The units of the photocurrent are given by $[e^4 v_F E_0^3/\pi \gamma_0 \hbar^2\Omega^2]$ for $J_1$ and $[e^4 v_F E_0^3/\pi \hbar^3\Omega^3]$ for $J_0$.
(f,g) Plots of $J_1$ as a function of the anisotropy $\eta$ for the five different light patterns shown in (a).
}
\label{fig_4}
\end{center}
\end{figure*}

Let us consider a two-frequency drive with elliptical polarization of ellipticity $\varepsilon_1,\varepsilon_2$,
\begin{align}\label{E_BCL}
    \vb*{E}^{(\Omega)} = E_0 \mqty(i \\ 0 \\ \varepsilon_1 ),\quad
    \vb*{E}^{(-2\Omega)} = E_0 e^{i\theta} \mqty(i \\ 0 \\ \varepsilon_2 ).
\end{align}
Each electric field corresponds to circularly polarized for $\varepsilon_i=\pm1$, linearly polarized for $\varepsilon_i=0$, and elliptically polarized for other values.
From Eqs.~(\ref{sigma_3D_inj},\ref{sigma_3D_shift}), we obtain the analytical expression of the photocurrent in the 3D Dirac systems as
\begin{align}\label{J_BCL1}
    J^x_1&=\frac{4e^4 |v_F|\Theta_0 E_0^3}{45\pi \gamma_0\hbar^2\Omega^2 |\eta|} \cos\theta \qty[
    -1+ \varepsilon_1(-3\varepsilon_1+4\varepsilon_2)\eta^2],
    \\ \label{J_BCL2}
    J^z_1&=\frac{4e^4 |v_F|\Theta_0 \eta^2 E_0^3}{45\pi \gamma_0\hbar^2\Omega^2 |\eta|} \sin\theta \qty[
    3\varepsilon_2 + \varepsilon_1(-4 + \varepsilon_1\varepsilon_2\eta^2)],
    \\ \label{J_BCL3}
    J^x_0&=\frac{-2e^4 |v_F|\Theta_0 E_0^3}{45\pi \hbar^3\Omega^3 |\eta|} \sin \theta \qty[
    23 + \varepsilon_1(9\varepsilon_1-32\varepsilon_2)\eta^2],
    \\ \label{J_BCL4}
    J^z_0&=\frac{-2e^4 |v_F|\Theta_0 \eta^2 E_0^3}{45\pi \hbar^3\Omega^3 |\eta|} \cos \theta \qty[
    9\varepsilon_2 +\varepsilon_1( -32+23\varepsilon_1\varepsilon_2 \eta^2)].
\end{align}
Similarly, from Eqs.~(\ref{cond_2Dmassless_inj},\ref{cond_2Dmassless_shi}), we obtain the photocurrent in the 2D Dirac system as 
\begin{align}
    J^x_1&=\frac{e^4 v_F^2 \Theta_0' E_0^3}{12 \gamma_0\hbar^2\Omega^3 |\eta|} \cos\theta \qty[
    -1+ \varepsilon_1(-5\varepsilon_1+6\varepsilon_2)\eta^2],
    \\
    J^y_1&=\frac{e^4 v_F^2 \Theta_0'\eta^2 E_0^3}{12 \gamma_0\hbar^2\Omega^3 |\eta|} \sin\theta \qty[
    5\varepsilon_2 + \varepsilon_1(-6 + \varepsilon_1\varepsilon_2\eta^2)],
    \\
    J^x_0&=\frac{-e^4 v_F^2 \Theta_0' E_0^3}{24 \hbar^3\Omega^4 |\eta|} \sin \theta \qty[
    17 - \varepsilon_1(11 \varepsilon_1 + 6 \varepsilon_2)\eta^2],
    \\ \label{J_BCL_2D}
    J^y_0&=\frac{-e^4 v_F^2 \Theta_0' \eta^2 E_0^3}{24 \hbar^3\Omega^4 |\eta|}  \cos \theta \qty[
    11\varepsilon_2 +\varepsilon_1(6 - 17\varepsilon_1\varepsilon_2 \eta^2)].
\end{align}
This shows that the direction of the two components of the photocurrent can be controlled with the relative phase $\theta$ of the two-frequency drive,
where the direction of the photocurrent $(J^x_0, J^z_0)$ or $(J^x_1, J^z_1)$ draws an ellipse with an ellipticity that depends on the parameters $\varepsilon_1,\varepsilon_2,\eta$ as the relative phase $\theta$ varies.
We note that the photocurrent vanishes in the case of the counter-rotating bicircular light, i.e., $(\varepsilon_1,\varepsilon_2)=(1,1)$, with $\eta\to1$ as $(J^x_1, J^z_1) \propto (1-\eta^2) (\cos\theta, \eta^2 \sin\theta)$ due to the three-fold rotational symmetry of the BCL \cite{Ikeda23}.
Figure~\ref{fig_4} shows the color plots of the ellipticity dependence of the four components of the photocurrent in 3D Dirac systems and the anisotropy dependence of the photocurrent for the five light patterns.
The ellipticity significantly affects the magnitude, direction, and material parameter dependence (here the anisotropy $\eta$) of the photocurrent.
Figures~\ref{fig_4}(b-e) showing the photocurrent components as a function of the ellipticities $\varepsilon_1$ and $\varepsilon_2$ indicates that the photocurrent in 3D Dirac systems is maximized for co-rotating bicircular light 
$(\varepsilon_1,\varepsilon_2)=\pm (1,-1)$. 

When applying a two-frequency drive with an inverted amplitude in the $x$ direction instead, $E^x \to -E^x$, only the photocurrent in the $x$ direction is found to be inverted as $(J^x,J^z)\to(-J^x,J^z)$.
This can be simply understood from which components of the conductivity are nonzero as listed in Eq.~\eqref{eight_cond}.
This phenomenon is similar to the second-order circular photogalvanic effect in time reversal symmetric systems derived from $\sigma_{\rm inj}^{zxy}$, but is different in that the current direction is changed (not reversed) in the BCL case.

Another interesting phenomenon is directional separation of shift and injection photocurrent responses.
Specifically, $J^x_1$ and $J^z_0$ are proportional to $\cos\theta$, whereas $J^z_1$ and $ J^x_0$ are proportional to $\sin\theta$, which is systematically determined by whether corresponding conductivity tensor components are real or purely imaginary.
For example, when $\theta=0$, a directional separation of photocurrent responses occurs such that the injection current and the shift current flows along the $x$ and $z$ directions, respectively.
Although such phenomena have been reported in magnetic systems~\cite{Ahn20}, dc response separation in non-magnetic systems has not been reported to the best of our knowledge.

\subsection{Linear polarization}
\begin{figure*}[t]
\begin{center}
\includegraphics[width=\linewidth]{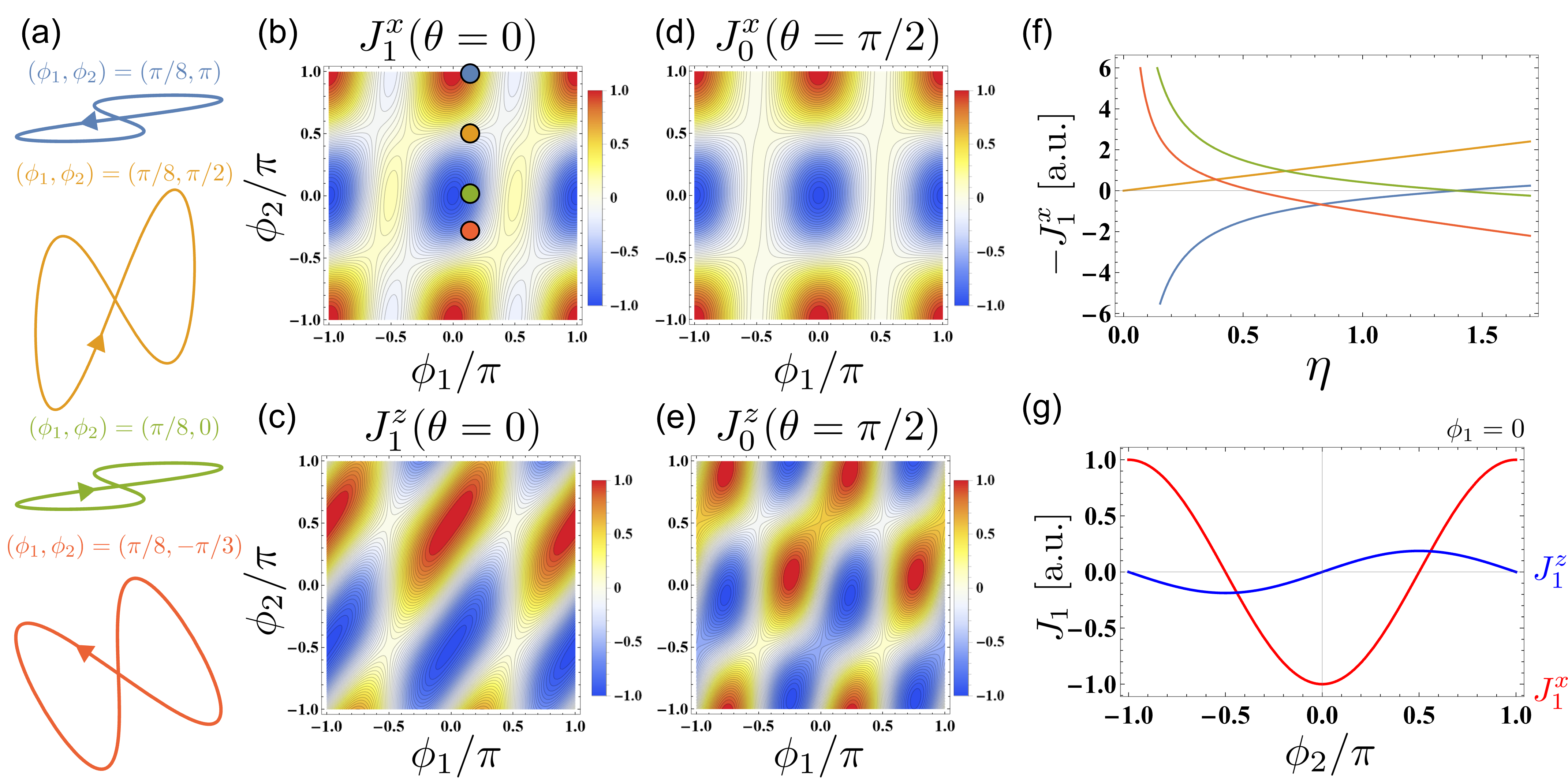}
\caption{
Photocurrent in 3D Dirac electron systems induced by linearly polarized two-frequency electric fields.
(a) Patterns of the linearly polarized two-frequency electric field with different polarization angles $(\phi_1,\phi_2)=(\pi/8,\pi),(\pi/8,\pi/2),(\pi/8,0),(\pi/8,-\pi/3)$ and $\theta$ is fixed to 0.
(b-e) Color plots of the photocurrent components as a function of the polarization angles $\phi_1$ and $\phi_2$. The dots in (b) correspond to the patterns of the two-frequency drive shown in (a) with matching colors. The calculations were performed with $\eta = 1/4$, $\theta=0$ in (b,c) and $\theta=\pi/2$ in (d,e).
(f) Plot of $J_1$ as a function of the anisotropy $\eta$ for the four different light patterns shown in (a).
(g) Plot of $J_1$ as a function of $\phi_2$  with fixing $\phi_1=0$, showing that the direction of the photocurrent can be controlled by tuning the polarization angle.
}
\label{fig_bilinear}
\end{center}
\end{figure*}

Next, we consider a two-frequency drive with two linearly polarized lights, which is called bilinear light,
\begin{align} \label{E_LPL}
    \vb*{E}^{(\Omega)} = iE_0 \mqty(\cos\phi_1 \\ 0 \\ \sin\phi_1),\quad
    \vb*{E}^{(-2\Omega)} = iE_0 e^{i\theta} \mqty(\cos\phi_2 \\ 0 \\ \sin\phi_2),
\end{align}
where $\phi_i$ represents the angle of polarization in the $x$-$z$ plane. 
In the bilinear light case, the photocurrent is given by
\begin{align}\label{bilinear_injx}
    J^x_1=&\frac{-4e^4 |v_F|\Theta_0 E_0^3}{45\pi \gamma_0\hbar^2\Omega^2 |\eta|} \cos\theta \Bigl[ 2\eta^2 \sin2\phi_1\sin\phi_2 \nonumber \\ 
    & +( \cos^2\phi_1 - 3\eta^2 \sin^2\phi_1
    ) \cos\phi_2
    \Bigr], \\   \label{bilinear_injz}
    J^z_1=&\frac{-4e^4 |v_F|\Theta_0 \eta^2 E_0^3}{45\pi \gamma_0\hbar^2\Omega^2 |\eta|} \cos\theta \Bigl[
    2\sin2\phi_1 \cos\phi_2 \nonumber \\
    &- (3\cos^2\phi_1 - \eta^2 \sin^2\phi_1 )\sin\phi_2 \Bigr], \\ 
    \label{bilinear_shix}
    J^x_0=&\frac{-2e^4 |v_F|\Theta_0 E_0^3}{45\pi \hbar^3\Omega^3 |\eta|} \sin \theta \Bigl[ 16\eta^2 \sin2\phi_1\sin\phi_2 \nonumber \\
    & +( 23\cos^2\phi_1 - 9 \eta^2 \sin^2\phi_1
    ) \cos\phi_2
    \Bigr], \\
    \label{bilinear_shiz}
    J^z_0=&\frac{-2e^4 |v_F|\Theta_0 \eta^2 E_0^3}{45\pi \hbar^3\Omega^3 |\eta|} \sin \theta \Bigl[ 16 \sin2\phi_1\cos\phi_2 \nonumber \\
    & -( 9\cos^2\phi_1 - 23 \eta^2 \sin^2\phi_1
    ) \sin\phi_2
    \Bigr],
\end{align}
for the 3D Dirac systems, 
and 
\begin{align}\label{2Dbilinear_injx}
    J^x_1=&\frac{-e^4 v_F^2 \Theta_0' E_0^3}{12 \gamma_0\hbar^2\Omega^3 |\eta|} \cos\theta \Bigl[ 3\eta^2 \sin2\phi_1\sin\phi_2 \nonumber \\ 
    & +( \cos^2\phi_1 - 5\eta^2 \sin^2\phi_1
    ) \cos\phi_2
    \Bigr], \\   \label{2Dbilinear_injz}
    J^z_1=&\frac{-e^4 v_F^2 \Theta_0' E_0^3}{12 \gamma_0\hbar^2\Omega^3 |\eta|} \cos\theta \Bigl[
    3\sin2\phi_1 \cos\phi_2 \nonumber \\
    &- (5\cos^2\phi_1 - \eta^2 \sin^2\phi_1 )\sin\phi_2 \Bigr], \\ 
    \label{2Dbilinear_shix}
    J^x_0=&\frac{-e^4 v_F^2 \Theta_0' E_0^3}{24 \hbar^3\Omega^4 |\eta|}
    \sin \theta \Bigl[ 3\eta^2 \sin2\phi_1\sin\phi_2 \nonumber \\
    & +( 17\cos^2\phi_1 + 11 \eta^2 \sin^2\phi_1
    ) \cos\phi_2
    \Bigr], \\
    \label{2Dbilinear_shiz}
    J^z_0=&\frac{-e^4 v_F^2 \Theta_0' E_0^3}{24 \hbar^3\Omega^4 |\eta|} \sin \theta \Bigl[ 3 \sin2\phi_1\cos\phi_2 \nonumber \\
    & +( 11\cos^2\phi_1 + 17 \eta^2 \sin^2\phi_1
    ) \sin\phi_2
    \Bigr],
\end{align}
for the 2D Dirac systems.
Unlike the bicircular light case, the phases of $E_x(t)$ and $E_z(t)$ are aligned for the bilinear light, and the relative phase dependence of the photocurrent is determined only by whether the dc conductivity is real or purely imaginary, which constrains that the injection current is proportional to $\sin\theta$ and the shift current to $\cos\theta$.
This indicates that the type of generated photocurrent can be selected by tuning the relative phase $\theta$ of the bilinear light.
Such phenomenon does not occur in the second-order dc generation by a monochromatic light.

Furthermore, the magnitude, direction, and $\eta$ dependence of the photocurrent can also be controlled by tuning the angle of polarization $\phi_1,\phi_2$ for the bilinear light.
Figures~\ref{fig_bilinear}(b-e) show the color plots of the polarization angle dependence of the photocurrent $J_1$ and $J_0$.
The four components of the photocurrent oscillate with respect to the angles $\phi_1$ and $\phi_2$.
For example, if $\phi_1$ is fixed to 0, the injection current draws an ellipse by varying $\phi_2$ as $(J^x_1,J^z_1) \propto (\cos\phi_2, -3\eta^2 \sin\phi_2)$ [see Fig.~\ref{fig_bilinear}(g)].
For other values of $\phi_1$, the trajectory of the current with respect to $\phi_2$ in the $x$-$z$ plane also shows various ellipses centered at the origin, as can be seen from the $\phi_2$ dependence of the photocurrent in  Eqs.~(\ref{bilinear_injx}-\ref{bilinear_shiz}).
Figure~\ref{fig_bilinear}(f) shows $J_1^x$ as a function of the anisotropy $\eta$ for several angle pairs, indicating qualitatively different anisotropy dependences of the photocurrent for different light patterns.
It can be seen that the polarization angle that maximizes the photocurrent changes radically with the value of the anisotropy of the Dirac cone.

We can consider other particular setups in which two linearly polarized lights are parallel or perpendicular to each other.
The photocurrent direction is given by
\begin{align}
    (J^x,J^z) &\propto [\cos^2\phi+\eta^2\sin^2\phi](\cos\phi, \eta^2\sin\phi), 
\end{align}
for the parallel orientation case ($\phi_1=\phi_2=\phi$),
and
\begin{align}
    (J^x_1,J^z_1) &\propto [1-7\eta^2 - (1-\eta^2)\cos2\phi](\cos\phi, \eta^2\sin\phi),
\end{align}
for the perpendicular polarization case,
$(\phi_1,\phi_2)=(\phi+\pi/2,\phi)$,
where the the current directions draw rather complex curves with respect to $\phi$ and their shapes depend on the anisotropy $\eta$.
In both cases, we find that the direction and magnitude of the photocurrent can be controlled by tuning the polarization angles.

\section{SYMMETRY CONSIDERATION AND DIMENSION ANALYSIS}\label{sec5}
In this section, we perform a dimensional analysis to study low-frequency divergence of photocurrent under the two-frequency drive.
We also perform symmetry consideration for the photocurrent induced by the two-frequency drive.

\subsection{Dimension analysis}\label{subsec_dim}
Since the Dirac electron system is described by the $k$ linear Hamiltonian and has a scale invariance,
the photocurrent response also shows a scale invariance under the change of the frequency $\Omega \to \lambda\Omega$
as
\begin{align}
    \ J(\lambda \Omega)=J(\Omega)/\lambda^\Delta,
\end{align}
with a scaling dimension $\Delta$.
A solution for this scale-invariant photocurrent takes the form
\begin{align}
    J(\Omega)\propto \Omega^{-\Delta}
\end{align}
leading to the expressions for $J_1$ and $J_0$ as
\begin{align}
    J_1& = C_1 \frac{e^a v_F^b E_0^3}{\hbar^c \gamma_0 \Omega^\Delta},\\
    J_0& = C_0 \frac{e^{a'} v_F^{b'} E_0^3}{\hbar^{c'} \Omega^{\Delta'}},
\end{align}
where $C_0,C_1$ are dimensionless constants including step functions for optical resonance conditions.
Since the dimension of the current density for 3D systems and 2D systems is $[\si{A/m^2}]$ and $[\si{A/m}]$, respectively, $a,b,c,\Delta$ are immediately obtained as
\begin{align}
    J_1& = C_1 \frac{e^4 v_F^{4-d} E_0^3}{\hbar^2 \gamma_0 \Omega^{5-d}},\\
    J_0& = C_0 \frac{e^{4} v_F^{4-d} E_0^3}{\hbar^{3} \Omega^{6-d}}.
\end{align}
Here, $d$ is the dimension of the system.
Note that the dimensionless constants $C_0,C_1$ are slightly different between 2D and 3D systems.

This dimensional analysis is found to be consistent with the results obtained from the diagrammatic calculations [Eqs.~(\ref{J_BCL1}-\ref{J_BCL_2D})].
Furthermore, considering the low-frequency limit, this shows that the photocurrent diverges as $J_1\propto \Omega^{-(5-d)}$ and $J_0\propto \Omega^{-(6-d)}$, which can be attributed to the gap-closing dispersion in Dirac/Weyl systems. Additionally, it is clear that the power of the low-frequency divergence is different for $J_1$ and $J_0$, and for the 3D and 2D systems.

For comparison, in the case of second-order photocurrents in inversion-broken Weyl systems, the scaling behavior is given by \cite{Ahn20}:
\begin{align}
J_1^{\text{2nd}} &\propto \tau \Omega^{-(3-d)}\\    
J_0^{\text{2nd}} &\propto \Omega^{-(4-d)}.
\end{align}
Thus, for third-order photocurrents, both the $\tau$-linear and $\tau$-independent contributions exhibit a stronger low-frequency divergence, with the power of the negative exponent increasing by two compared to the second-order case.

\subsection{Symmetry consideration}
Next, we consider how the photocurrent change under time reversal operation $\mathcal{T}$.
The general third-order response of the current density $J$ is defined as
\begin{align}
    J(\Omega_{\rm tot})=\sigma^{\mu\alpha\beta\gamma}(\Omega_{\rm tot};\Omega_1,\Omega_2,\Omega_3)E_\alpha (\Omega_1)E_\beta (\Omega_2)E_\gamma (\Omega_3).
\end{align}
Under the time reversal operation $\mathcal{T}$, the current density transforms as
\begin{align}
    J(\Omega_{\rm tot}) \xrightarrow{\mathcal{T}}  -J(-\Omega_{\rm tot}),
\end{align}
which leads that the third-order photocurrent satisfies
\begin{align}
    J_1 \xrightarrow{\mathcal{T}} J_1, \quad
    J_0 \xrightarrow{\mathcal{T}} -J_0
\end{align}
for $J_1 \propto 1/\gamma_0$ and $J_0 \propto O((\gamma_0)^0)$,
since $\Omega_{\rm tot}=3i\gamma_0$ in our setup.
On the other hand, the electrical field is converted under $\mathcal{T}$ as
\begin{align}
    E(\Omega_i) \xrightarrow{\mathcal{T}}  E(-\Omega_i)=E^*(\Omega_i).
\end{align}
In the case of circularly polarized two-frequency electric fields [Eq.~(\ref{E_BCL})], the complex conjugation corresponds to $(E_0, \theta, \varepsilon)\to (-E_0, -\theta, -\varepsilon)$.
Therefore, we can conclude that $J_1$ is even and $J_0$ is odd for the transformation $(\theta, \varepsilon)\to (-\theta, -\varepsilon)$.
This is consistent with the analytical expressions in Eqs.~(\ref{J_BCL1}-\ref{J_BCL4}). Furthermore, Figs. \ref{fig_4}(b-e) allow for the verification of the sign of the current with respect to the inversion of the ellipticity.
In the case of linearly polarized two-frequency electric fields [Eq.~(\ref{E_LPL})], the complex conjugation corresponds to $(E_0, \theta)\to (-E_0, -\theta)$, leading that $J_1$ is $\theta$-even and $J_0$ is $\theta$-odd. This also agrees with the results in Eqs.~(\ref{bilinear_injx}-\ref{bilinear_shiz}).

\section{Discussion}\label{sec6}
As discussed in Sec.~\ref{sec_circular}
, utilizing a circularly polarized two-frequency drive enables the directional separation of shift and injection photocurrents, allowing for the separate detection of these two types of photocurrents. 
From Eqs.~(\ref{J_BCL2},\ref{J_BCL3}) with $(\varepsilon_1,\varepsilon_2)=(1,-1)$, their ratio is given by
\begin{align}
    \frac{J_1^z}{J_0^x} = \frac{\eta^2(7+\eta^2)}{23+41\eta^2}\frac{\hbar\Omega}{\gamma_0}
    = \frac{\eta^2(7+\eta^2)}{23+41\eta^2}\tau \Omega
\end{align}
and determined by the ratio of the input frequency $\Omega$ to the system's relaxation rate $\gamma_0$.
This is because $J_1$ and $J_0$ have different powers for the low-frequency divergence.
Inserting the typical parameters, $\hbar \Omega=0.1 \, \si{eV}$ and $\tau=\hbar/\gamma_0=1\,$ps, we obtain
 \begin{align}
    \frac{J_1^z}{J_0^x} \sim \frac{15\eta^2(7+\eta^2)}{23+41\eta^2}
\end{align}
For the typical example of a Dirac semimetal, Cd$_3$As$_2$, where $\eta \simeq 1/4$, it can be seen that the magnitude of $J_1$ and $J_0$ are comparable as $J_1^z/J_0^x\sim0.26$.

In 3D Dirac systems, the estimated magnitude of the third-order photocurrent density is approximately $J_1^{\rm 3D}\sim 10^6\, \si{A/m^2}$, adopting realistic parameters, $\hbar \Omega=0.1 \, \si{eV}$, $E_0=1 \,$kV/cm, $v_{\rm F}=10^6\,$m/s, and $\tau=\hbar/\gamma_0=1\,$ps. For a sample with a width of $L=100\,$\si{\micro \meter} and a penetration depth of $\delta=1\,$\si{\micro \meter}, the magnitude of the photocurrent is estimated to be $I^{\rm 3D}=J_1^{\rm 3D} L \delta\sim 100\,$\si{\micro \ampere}, indicating that a large photocurrent is expected under a two frequency drive in the mid-infrared region. The substantial enhancement in $J$ is caused by the low-frequency divergence of the conductivities due to the gapless nature of Dirac semimetals, as suggested in sections \ref{sec4} and \ref{subsec_dim}.
In contrast, for a two-dimensional (2D) Dirac system like graphene, the current density is given by
\begin{align}
    J_1^{\rm 2D} \sim J_1^{\rm 3D} \frac{v_F}{\Omega}.
\end{align}
Hence, the photocurrent is estimated to be $I^{\rm 2D}=J_1^{\rm 2D} L\sim 1\,$\si{\micro \ampere}, which is also large enough to be observed experimentally.
These large photocurrent responses in gapless systems using a two-frequency drive provide a new venue to the dynamic control of photocurrents with possible applications in optoelectronic devices in the mid-infrared or terahertz range.

\section{Conclusion}
In conclusion, our work provides a systematic analysis of third-order photocurrent generation in Dirac and Weyl systems under two-frequency light irradiation. We emphasize several key novelties: first, we thoroughly investigate both $\tau$-linear and $\tau$-independent contributions to the photocurrent, providing a more complete picture of the BPVE in these systems. Notably, we reveal that the gapless dispersion at Dirac and Weyl points leads to a significant low-frequency divergence in the photocurrent. This divergence is stronger than that of second-order photocurrents in inversion-broken Weyl systems, with the frequency's negative exponent increasing by two for both $\tau$-linear and $\tau$-independent terms. Second, we successfully derive analytical expressions for the conductivity in 2D and 3D Dirac systems, including massive/massless Dirac and tilted Dirac systems. Additionally, we demonstrate that by altering the polarization of incident laser light, it is possible to dynamically control the photocurrent's direction and type, enabling tunable optoelectronic functionality. These findings suggest that Dirac and Weyl materials hold great potential for future applications in controllable optoelectronic devices, where both the magnitude and direction of photocurrents can be engineered through light manipulation.

\acknowledgments
This work was supported by 
JSPS KAKENHI Grants
No. 24KJ0731 (Y.I.),
No. 20K14407 (S.K.), 
No. 23K25816, No. 23K17665, and No. 24H02231 (T.M.), 
and
JST CREST (Grant No. JPMJCR19T3) (S.K., T.M.).

\bibliographystyle{apsrev4-1}
\bibliography{reference.bib}
\clearpage

\onecolumngrid
\appendix

\section{Diagrammatic approach for photocurrent}\label{app}

In this appendix, we briefly explain the diagrammatic approach to nonlinear optical responses and the application for third-order dc generation in Dirac and Weyl systems.
Let us consider an unperturbed Hamiltonian written as
\begin{align}
    H_0 =\sum_a
    \int [dk] \epsilon_a c^\dagger_{\vb*{k}a}c_{\vb*{k}a}
\end{align}
where $a$ is a band index
and $c^\dagger_{\vb*{k}a} (c_{\vb*{k}a})$ is the fermionic creation (annihilation) operator and we use a shorthand notation $[\dd\vb*{k}]\equiv \dd\vb*{k}/(2\pi)^d$ with the spatial dimension $d$.

Following Refs.~\cite{Parker19,Passos2018,YuTzu2024}, the Hamiltonian under an electric field can be expanded as
\begin{align}
    H(t)
    =H_0+
    \sum_{n=1}^\infty \frac{1}{n!}\qty(\frac{e}{\hbar})^n
    (\vb*{A}(t)\vdot \vb*{D})^n H_0\equiv H_0+ V(t) ,
\end{align}
where $\vb*{A}(t)$ is the vector potential of the electric field.
Here, $\vb*{D}$ is the covariant derivative defined as
\begin{align}
    (D^\alpha \mathcal{O})_{ab}=
    \pdv{\mathcal{O}_{ab}}{k_\alpha}
    -i [\mathcal{A}^\alpha,\mathcal{O}]_{ab},
\end{align}
where the Berry connection matrix $\mathcal{A}^\alpha$ is given by $\mathcal{A}^\alpha_{ab}=i\bra{a}\partial_{k_\alpha} \ket{b}$ with the Bloch wave function $H_0 \ket{a} = \epsilon_a \ket{a}$.
Using the Fourier transformation of the vector potential
\begin{align}
    A^{\alpha_l}(t)=
    \int d\Omega_l
    e^{-i\Omega_l t}\frac{E^{\alpha_l}(\Omega_l)}{i\Omega_l},
\end{align}
the perturbation part $V(t)$ can be written as
\begin{align}
    V(t)=
    \sum_{n=1}^\infty \frac{1}{n!}\qty(\frac{e}{\hbar})^n
    \qty[
    \prod_{l=1}^n  \int d\Omega_l
    e^{-i\Omega_l t}\frac{E^{\alpha_l}(\Omega_l)}{i\Omega_l}
    ]
    h^{\alpha_1\cdots \alpha_n}
\end{align}
with the higher derivative of the unperturbed Hamiltonian $h^{\alpha_1\cdots \alpha_n}=D^{\alpha_1}\cdots D^{\alpha_n}H_0$. Note that the summation about $\alpha_l\in \{x,y,z\}$ is taken implicitly.
The current density can be calculated from the expectation value of the current operator in the form of a path integral as
\begin{align}
   \expval{J^\mu(t)} &= \frac{1}{Z}\int \mathcal{D}[c,c^\dagger] (-e v_E^\mu(t)/\hbar)\exp\qty(-i\int dt' H(t')), \\
   Z&=\int \mathcal{D}[c,c^\dagger]
   \exp\qty(-i\int dt H(t)), \\
   v_E^\mu(t) &= 
   D^\mu H(t)= D^\mu [H_0+V(t)]=
   \sum_{n=0}^\infty \frac{1}{n!}\qty(\frac{e}{\hbar})^n
    \qty[
    \prod_{l=1}^n  \int d\Omega_l
    e^{-i\Omega_l t}\frac{E^{\alpha_l}(\Omega_l)}{i\Omega_l}
    ]
    h^{\mu\alpha_1\cdots \alpha_n}.
\end{align}
Finally, we can obtain the frequency domain optical conductivity in the form of a functional derivative as
\begin{align}\label{J_mu}
    \sigma^{\mu\alpha_1\cdots\alpha_n}
    (\Omega;\Omega_1,\cdots,\Omega_n) &=
    \frac{\delta}{\delta E^{\alpha_1}(\Omega_1)}\cdots
    \frac{\delta}{\delta E^{\alpha_n}(\Omega_n)} \expval{J^\mu (\Omega)}\\
    &= \int\frac{dt}{2\pi}e^{i\Omega t} \prod_{l=1}^n \qty[
    \int \frac{dt_l}{2\pi}e^{-i\Omega_l t_l}\frac{\delta}{\delta E^{\alpha_1}(t_l)}] \expval{J^\mu(t)}\\
    &=\int\frac{dt}{2\pi}e^{i\Omega t} \prod_{l=1}^n \qty[
    \int \frac{dt_l}{2\pi}e^{-i\Omega_l t_l} ]\sigma^{\mu\alpha_1\cdots\alpha_n}
    (t;t_1,\cdots,t_n).
\end{align}
Therefore, the conductivity in the time domain should be calculated, which can be obtained from the functional derivative of Eq~\eqref{J_mu}.
The electric field is incorporated in both $v^\mu_E(t)$ and $V(t)$, and hence there are $2^n$ contributions to the $n$th-order conductivity, depending on from which the functional derivative is taken.
See Refs.~\cite{Parker19,YuTzu2024} to know detailed expression of the first-order and second-order conductivities.

Previous studies have derived that such calculation of the optical conductivity by path integral can be performed systematically by the diagram method.
For $n$-th order optical conductivities, it is sufficient to consider $2^n$ diagrams that satisfy the following conditions: (i) There are $n+1$ external photons. (ii) Diagram forms a loop. (iii) One vertex represents the output current $J^\mu$. (iv) The value of edges and vertices in diagrams are shown in Table~\ref{tb1}.
For the derivation of these rules, see Refs.~\cite{Parker19,YuTzu2024}.
Using this technique, we can systematically calculate general optical conductivities including third-order dc conductivities, which are focused on in this paper.

\begin{table}[htbp]
  \centering
  \begin{tabular}{C{4cm}{}{m} C{4cm}{}{m} C{4cm}{}{m}}
Component & Diagram & Value \nextRow
\hline\hline Propagator & \includegraphics[width=2.5cm]{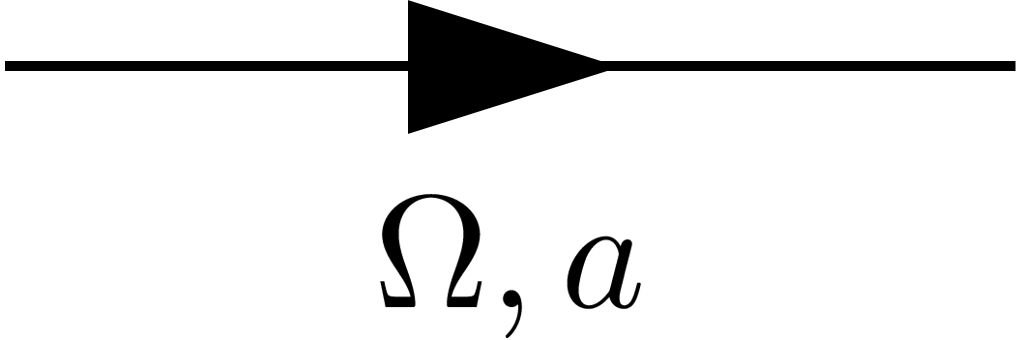} &  
    $\displaystyle
    G_a(\Omega)=\frac{1}{\Omega-\epsilon_a+i\gamma_0}
    $   \nextRow \addlinespace[0.5em]
    \hline 
    $n$-photon input vertex & \includegraphics[width=3cm]{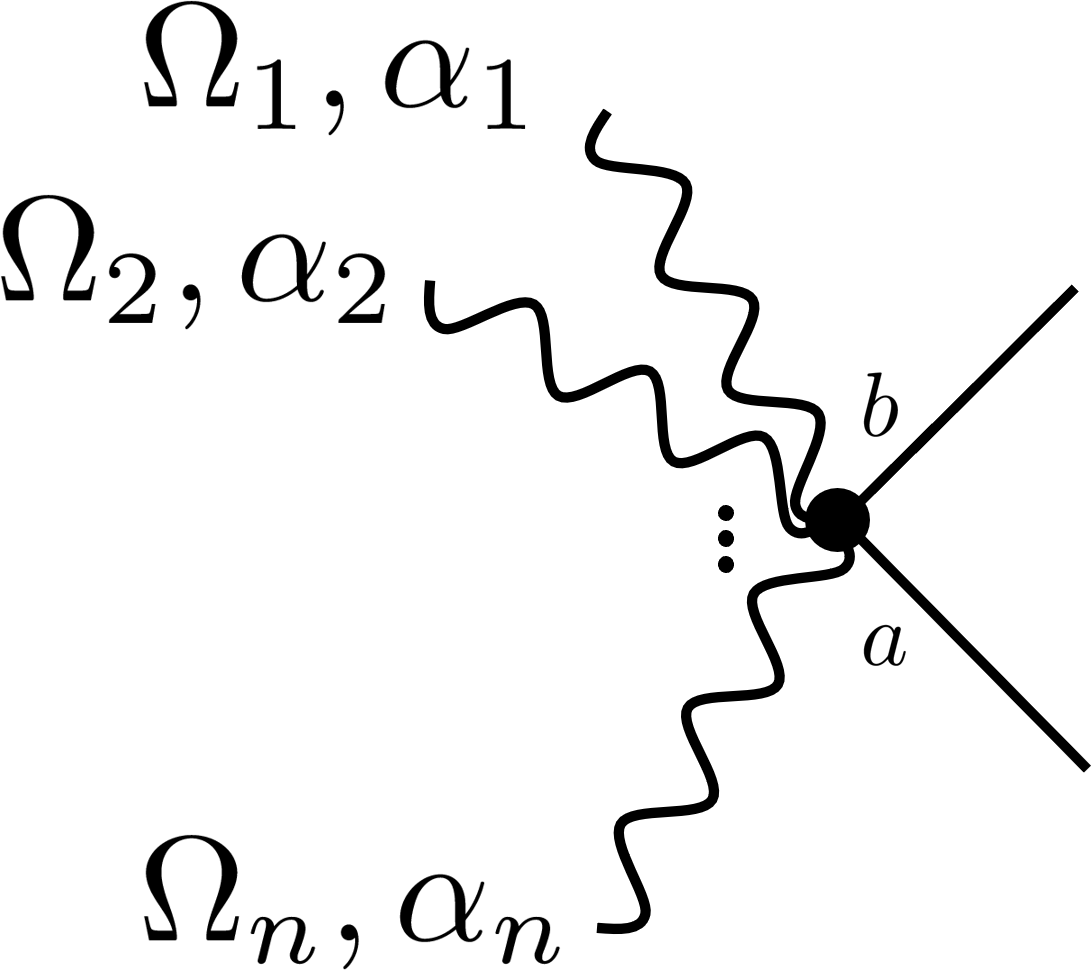} & $
    \displaystyle \frac{1}{n!}\prod_{i=1}^n \qty[\frac{ie}{\hbar\Omega_i}]h^{\alpha_1\alpha_1\cdots\alpha_n}_{ba}$
    \nextRow \addlinespace[0.5em]
    \hline 
    $(n+1)$-photon output vertex & \includegraphics[width=3cm]{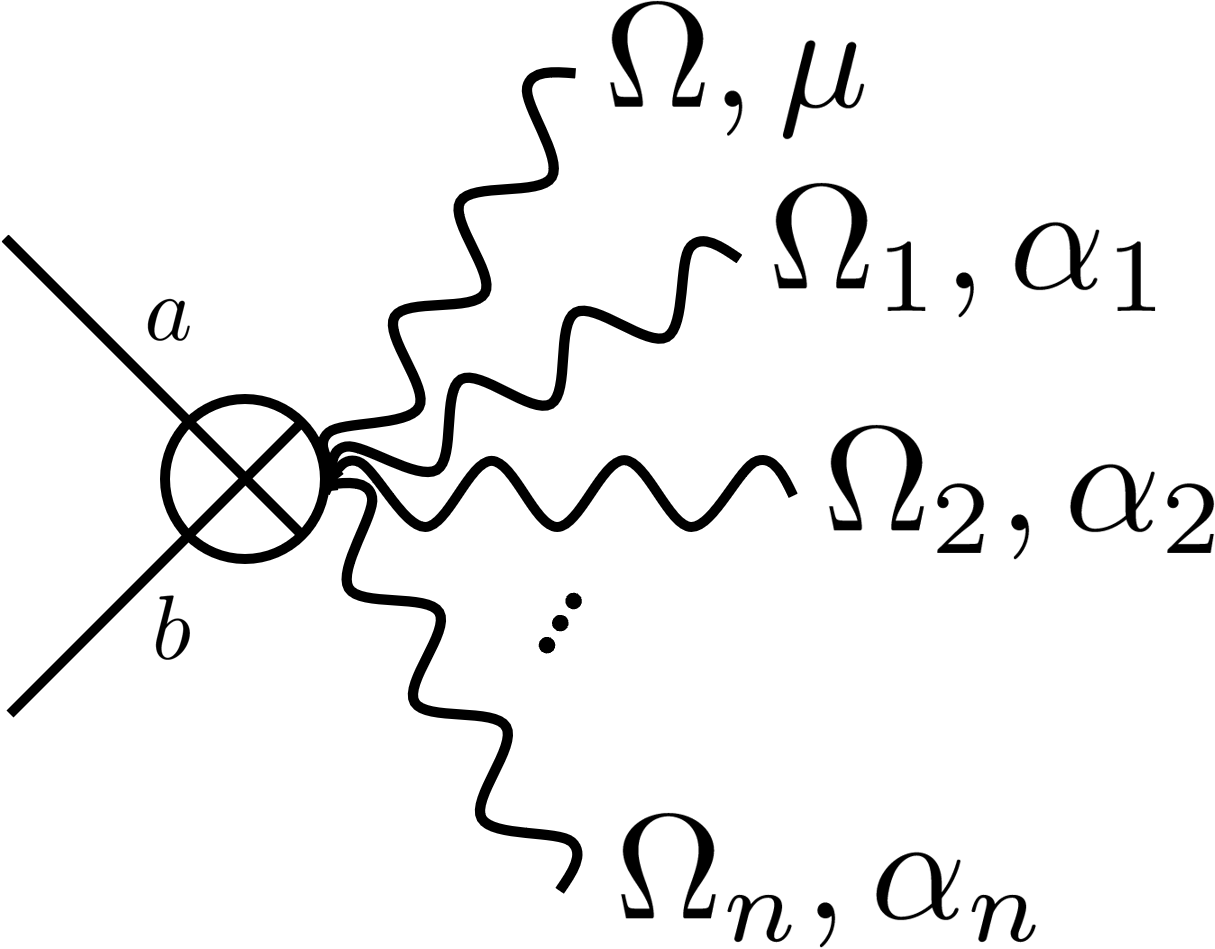} & 
    $\displaystyle \frac{1}{n!}\frac{e}{\hbar}
    \prod_{i=1}^n
    \qty[\frac{ie}{\hbar\Omega_i}]
    h^{\mu\alpha_1\alpha_1\cdots\alpha_n}_{ba}$
    \nextRow \addlinespace[0.5em]
\hline\hline
\end{tabular}
  \caption{The Feynman rules for the diagram components for the optical conductivity.}
  \label{tb1}
\end{table}

General third-order nonlinear optical conductivities can be written with eight diagrams as follows:

\begin{align}\label{eq_diagram_all}
\begin{split}
  \sigma^{\mu\alpha\beta\gamma} &= \; \raisebox{-0.4\height}{\includegraphics[height=1.5cm]{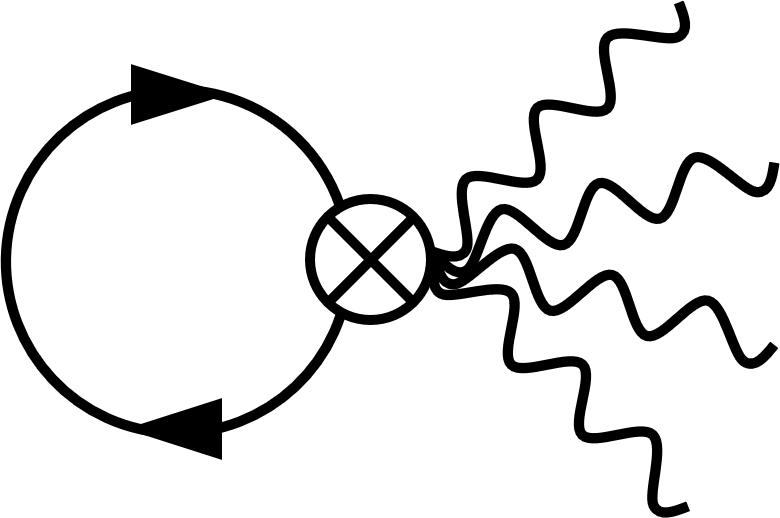}} \; + \;  
       \raisebox{-0.4\height}{\includegraphics[height=1.5cm]{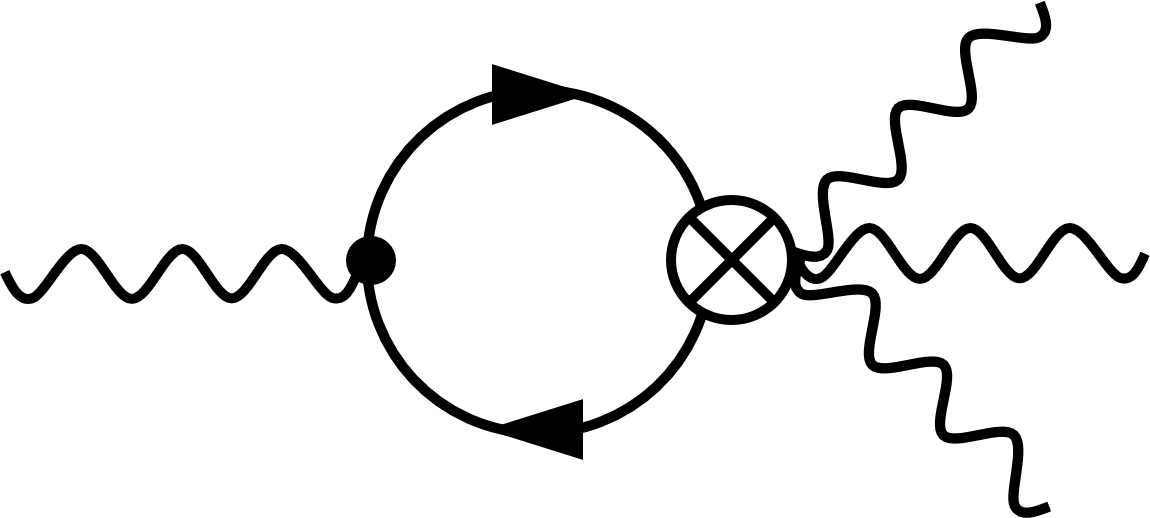}}\; + \; 
       \raisebox{-0.5\height}{\includegraphics[height=1.2cm]{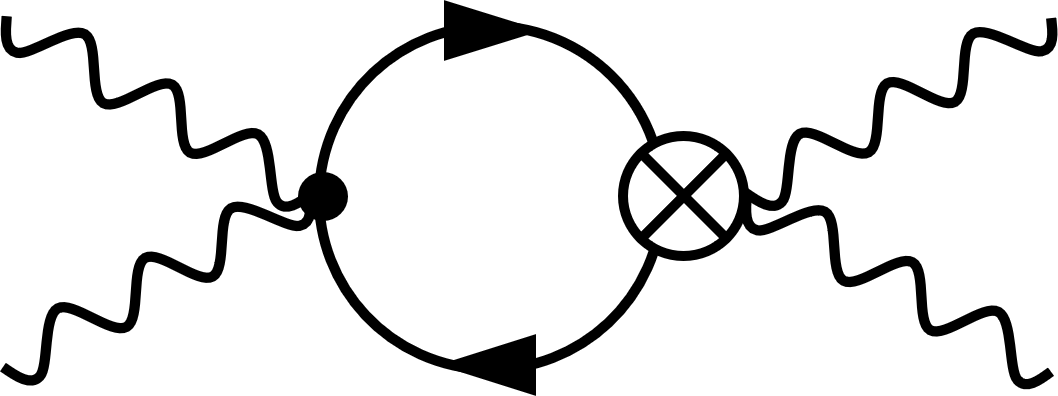}}\; + \; 
       \raisebox{-0.45\height}{\includegraphics[height=1.5cm]{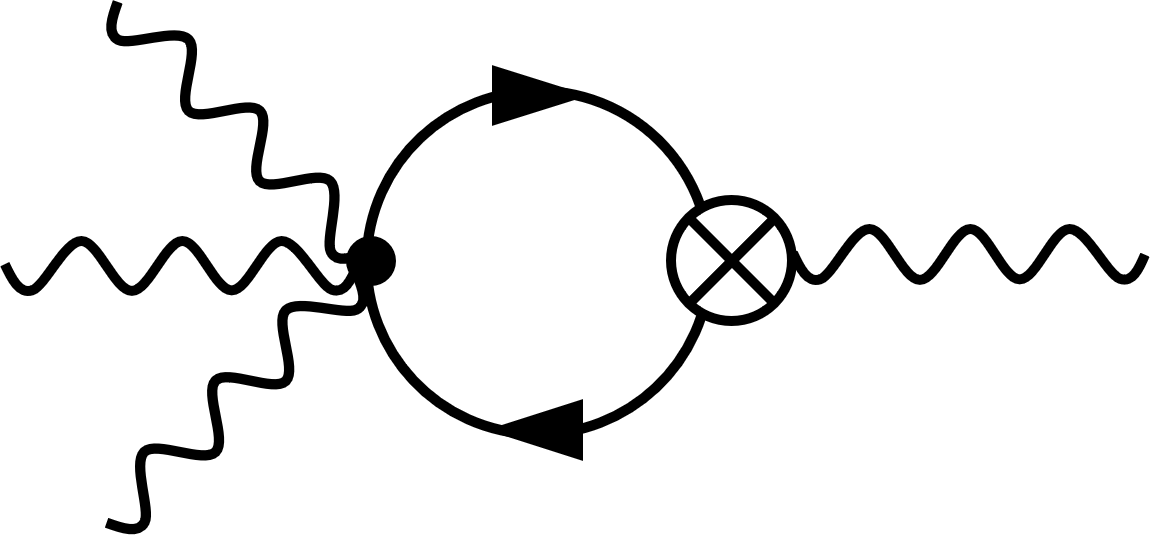}} \\
    &+ \; \raisebox{-0.4\height}{\includegraphics[height=1.3cm]{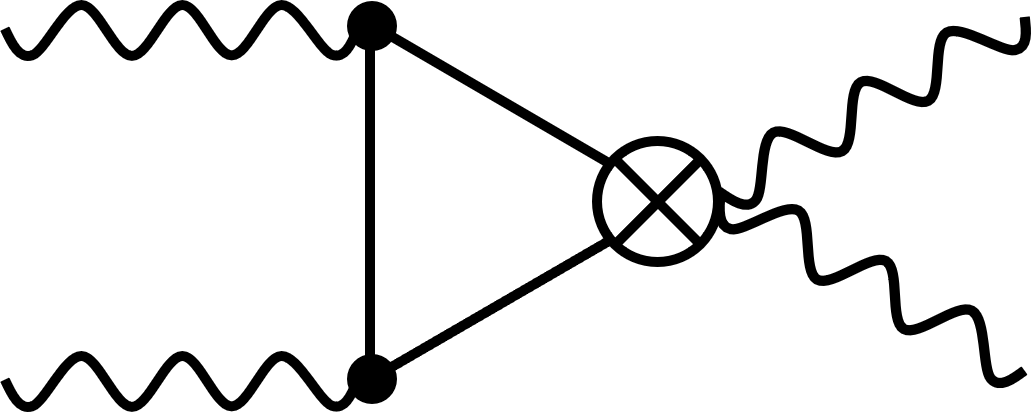}} \; + \;
       \raisebox{-0.4\height}{\includegraphics[height=1.7cm]{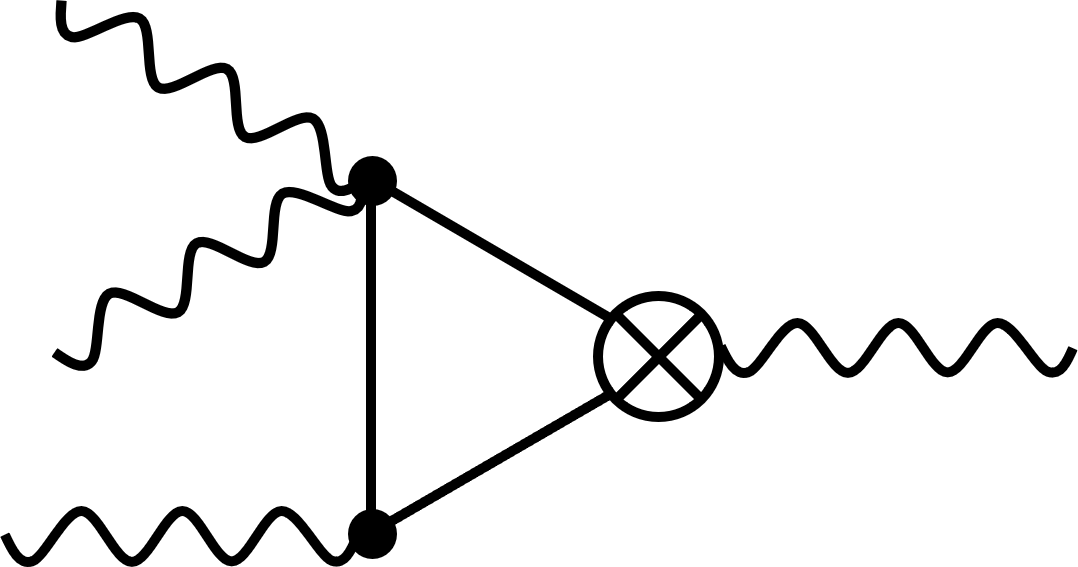}}\; + \;
       \raisebox{-0.6\height}{\includegraphics[height=1.7cm]{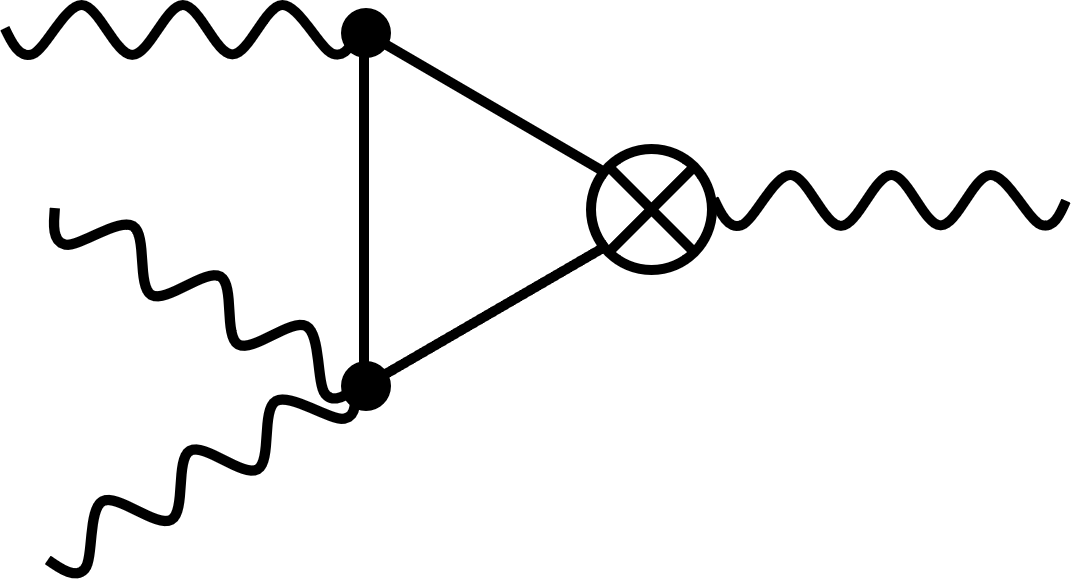}}\; + \;
       \raisebox{-0.4\height}{\includegraphics[height=1.5cm]{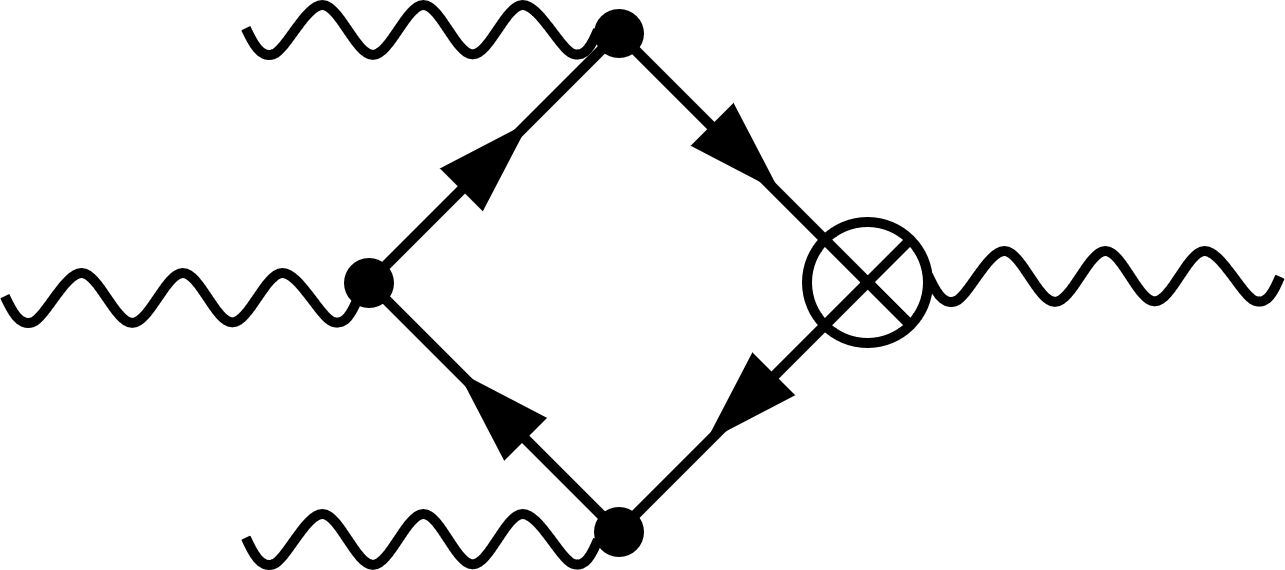}}
\end{split}
\end{align}
The contribution from each diagram can be calculated by the Feynman rule described above.
Although Eq.~(\ref{eq_diagram_all}) involves many terms and is highly complicated, it becomes significantly simpler in the Dirac/Weyl systems targeted in this paper. For instance, in the case of a three-dimensional system, the Hamiltonian consists only of $k$-linear terms as
\begin{align}\label{app_ham}
    H^{\rm 3D}= \hbar v_F (k_x \sigma_x + k_y \sigma_y + \eta k_z \sigma_z).
\end{align}
Furthermore, the covariant derivative $h^\alpha$ reduces to 
$h^\alpha_{ab}=v^\alpha_{ab}\equiv
(\partial_{k_\alpha}H_0(\vb*{k}))_{ab}$,
as shown in the following: (i) when $a=b$,
\begin{align}
    h^\alpha_{aa}&=\pdv{(H_0)_{aa}}{k_\alpha}
    -i [\mathcal{A}^\alpha,H_0]_{aa}
    =\pdv{\epsilon_a}{k_\alpha}
    -i \sum_b [ \mathcal{A}^\alpha_{ab} (H_0)_{ba}
    -(H_0)_{ab}\mathcal{A}^\alpha_{ba}]\\
    &=\pdv{\epsilon_a}{k_\alpha} = v^\alpha_{aa},
\end{align}
with $(H_0)_{ab}=\epsilon_a \delta_{ab}$ and (ii) when $a\neq b$,
\begin{align}
     h^\alpha_{ab}&=\pdv{(H_0)_{ab}}{k_\alpha}
    -i [\mathcal{A}^\alpha,H_0]_{ab}
    =-i\sum_c[\mathcal{A}^\alpha_{ac} (H_0)_{cb}
    -(H_0)_{ac}\mathcal{A}^\alpha_{cb}]\\
    &= i \epsilon_{ab}\mathcal{A}^\alpha_{ab}=v^\alpha_{ab}.
\end{align}
Consequently, higher-order covariant derivatives (i.e., $h^{\alpha\beta}, h^{\alpha\beta\gamma}, h^{\mu\alpha\beta\gamma}$) vanish in the Dirac/Weyl systems since $v^\alpha_{ab}$ is constant. Therefore, third-order optical responses in systems composed only of $k$-linear terms, as in Eq.~(\ref{app_ham}), can be fully described by the box diagram containing only the one-photon vertex (the last diagram in Eq.~(\ref{eq_diagram_all})).
As described in the main text, by setting the input frequencies to $\{\Omega_1,\Omega_2,\Omega_3 \}=\{\Omega,\Omega,-2\Omega \}$, we obtain the general expression for the two-frequency driving induced photocurrent as
\begin{align}\label{app_sigma}
    \sigma^{\mu\alpha\beta\gamma}(0;\Omega,\Omega,-2\Omega)&=
    \frac{-ie^4}{2\hbar^4\Omega^3}\mathcal{S}\sum_{abcd}\int[\dd\vb*{k}]v_{ba}^\alpha v_{cb}^\beta v_{dc}^\gamma v_{ad}^\mu  I_4(\hbar\Omega+i\gamma_0,\hbar\Omega+i\gamma_0,-2\hbar\Omega+i\gamma_0),
\end{align}
where $\mathcal{S}$ denotes the summation for all possible permutations of input photons $(\alpha,\Omega),(\beta,\Omega),(\gamma,-2\Omega)$.
The frequency integral in the box diagram, which we call $I_4$, is performed for imaginary-time Green's functions as \cite{Parker19}
\begin{align}
\begin{split}\label{app_I4}
I_4(i\Omega_1,i\Omega_2,i\Omega_3)&=\int \frac{\dd\omega}{2\pi}    G_a(i\omega)G_b(i\omega+i\Omega_1)G_c(i\omega+i\Omega_1+i\Omega_2)G_d(i\omega+i\Omega_1+i\Omega_2+i\Omega_3)\\
    &=\frac{f(\epsilon_a)}{(\epsilon_{ab}+i\Omega_1)(\epsilon_{ac}+i\Omega_{12})(\epsilon_{ad}+i\Omega_{123})}
        +\frac{f(\epsilon_b)}{(\epsilon_{ba}-i\Omega_1)(\epsilon_{bc}+i\Omega_{2})(\epsilon_{bd}+i\Omega_{23})}\\
        &+\frac{f(\epsilon_c)}{(\epsilon_{ca}-i\Omega_{12})(\epsilon_{cb}-i\Omega_{2})(\epsilon_{cd}+i\Omega_{3})}
        +\frac{f(\epsilon_d)}{(\epsilon_{da}-i\Omega_{123})(\epsilon_{db}-i\Omega_{23})(\epsilon_{dc}-i\Omega_{3})}. 
\end{split}    
\end{align}
After analytic continuation of Matsubara frequencies,
\begin{align}
    i\Omega_1 &\to \hbar\Omega+i\gamma_0, &
    i\Omega_2 &\to \hbar\Omega+i\gamma_0, &
    i\Omega_3 &\to -2\hbar\Omega+i\gamma_0,
\end{align}
in the above expression and $k$-integration in Eq.~(\ref{app_sigma}), we can obtain the analytical expression of the third-order dc conductivities in Dirac and Weyl systems as described in the main text.

\end{document}